\documentclass[aps, prb, reprint, groupedaddress]{revtex4-2}

\usepackage{amsmath, amssymb}
\usepackage{braket}
\usepackage{multirow}
\usepackage{vtable}
\usepackage{algorithmic}
\usepackage{mathtools}
\usepackage{amsfonts}
\usepackage{extarrows}

\usepackage{graphicx}
\usepackage{color}
\usepackage{bm}
\usepackage{hyperref}

\usepackage{algorithmic}
\usepackage{algorithm}

\allowdisplaybreaks[1]

\usepackage[normalem]{ulem}
\usepackage{comment}

\begin{document}
\title{Photoemission orbital tomography based on tight-binding approach: method and application to $\pi$-conjugated molecules}

\author{Misa Nozaki}
\email{6d697361@gmail.com}
\affiliation{Graduate School of Science and Engineering, Chiba University, Yayoi-cho 1-33, Inage, Chiba 263-8522, Japan}
\author{Takehisa Konishi}
\email{konishi@faculty.chiba-u.jp}
\affiliation{Graduate School of Science and Engineering, Chiba University, Yayoi-cho 1-33, Inage, Chiba 263-8522, Japan}
\date{\today}


\begin{abstract} 
Conventional photoemission orbital tomography based on Fourier iterative method enables us to extract a projected two-dimensional (2D) molecular orbital from a 2D photoelectron momentum map (PMM) of planar $\pi$-conjugated molecules in a single-orientation system, while not in a multi-orientation system. In this work, we demonstrate photoemission orbital tomography for $\pi$-conjugated molecules with a tight-binding ansatz using atomic orbital basis. 
We analyze 2D PMMs of single-orientation pentacene/Ag(110) and multi-orientation 3,4,9,10-perylenetetracarboxylic dianhydride/Ag(110) and reproduce their three-dimensional highest occupied molecular orbitals. We demonstrate that the PhaseLift algorithm can be used to analyze PMM including experimental or theoretical uncertainties. With the 2D PMM for pentacene, we simultaneously optimized the structure and the molecular orbital. The present approach enables us to extract the three-dimensional orbitals and structures of existing materials.
\end{abstract}

\maketitle

\section{Introduction}
\label{sec:intro}
Angle-resolved photoemission spectroscopy (ARPES) is an indispensable experimental method to obtain information on the electronic states. ARPES has been intensively used to address the electronic states of organic thin films adsorbed on solid surfaces \cite{NKU20, PUS09, PRU11, ULR14, IKS22}.
The recently developed time- and angle-resolved photoemission spectroscopy \cite{KXD20, ZPJ22} made it possible to record real-time non-equilibrium electronic processes in organic thin films \cite{BRM22}.

Puschnig and others \cite{PUS09, LUR14} developed a method to reproduce molecular orbitals from photoelectron momentum map (PMM) by recovering the phase of the Fourier-transformed orbital. This method is called photoemission orbital tomography (POT). PMM is proportional to the absolute square of the Fourier-transformed molecular orbital within the framework of the plane-wave final state approximation and does not contain information on the Fourier phases. The currently proposed POT \cite{LUR14, OLU15, KLO16, KLZ18, JAN20} uses the Fourier iterative method \cite{GS72, Fie82} and its variants \cite{MHC03,JAN20} as phase retrieval algorithms. These POTs have been mainly used to reconstruct two-dimensional (2D) projections of molecular orbitals from 2D PMMs of single-orientated planar $\pi$-conjugated molecules adsorbed on metallic substrates \cite{LUR14, OLU15, KLO16, KLZ18, JAN20}.
L\"uftner {\it et al}. also accomplished 3D reconstruction of frontier orbitals of single-orientation 3,4,9,10-perylenetetracarboxylic dianhydride (PTCDA) by varying the photon energy \cite{WLU15}. One should, however, note that preparing such a 3D PMM requires prudent calibration of the intensities of light for different energies \cite{WLU15, GMG19}.

For microscopic insights into various organic semiconductors, further developments are desired. 
First, reconstructing a reliable 3D molecular orbital from a single set of 2D PMM is desired. 
From a 2D PMM, we can reproduce at most a 2D projection of molecular orbital with the conventional POT based on the Fourier iterative method.
Kliuev {\it et al}. discussed the possibility to reconstruct a full 3D orbital from a 2D PMM by using an ansatz, but did not show concrete examples \cite{KLZ18}. 
Second, even in epitaxial molecular films, it is quite common to find not one but several symmetry-related molecular orientations. 
Therefore, it is important to extend POT to multi-orientation systems.
The conventional POT based on the Fourier iterative method cannot treat incoherent superposition of photoelectron intensities from molecules with various orientations \cite{KZJ19}.

In this work, we propose a POT based on a tight-binding picture and apply it to reproducing molecular orbitals of $\pi$-conjugated molecules on metal substrate.  
We formulate the POT as a fitting of experimental PMM to theoretical photoelectron intensity from molecular orbitals.
By describing the molecular orbitals with localized atomic orbitals centered on fixed atomic positions, the problem becomes simpler. 
To solve the problem, we use the least squares fitting and the PhaseLift algorithm \cite{CSV13, CES15, footnote}. We reproduce the 3D highest occupied molecular orbitals (HOMOs) of single-orientation pentacene and multi-orientation PTCDA [Fig. \ref{fig:geometry}(a) and \ref{fig:geometry}(c)] from their 2D PMMs. In the case of pentacene, we also remove the constraint of fixed atomic positions by simultaneously optimizing the molecular orbital coefficients and the molecular structure with the experimental 2D PMM.

\section{Theory}
\subsection{Photoelectron intensity of molecules}
\label{Sec:intensity}
In the plane wave final state approximation, we can describe the photoelectron intensity from an adsorbed molecule on metal substrate as 
\begin{eqnarray}
W_{{\bf k}}(\ket{\psi_i})
&\propto& | \langle {\bf k} | {\bm \varepsilon} \cdot {\bf p} | \psi_i \rangle |^2 \delta\left(h\nu - E_i - E_k-\Phi \right)
\nonumber\\
&\propto& |{\bm \varepsilon} \cdot {\bf k}|^2 |\braket{{\bf k}|\psi_i}|^2 \delta\left(h\nu - E_i - E_k-\Phi\right),
\label{eq:W}
\end{eqnarray} 
using Fermi's golden rule.
Here $W_{{\bf k}}(\ket{\psi_i})$ denotes the transition rate from an initial state $\ket{\psi_i}$ with binding energy $E_i$ relative to the Fermi energy of the substrate to a plane wave final state $\ket{{\bf k}}$ with kinetic energy $E_k$. 
$h\nu$ and ${\bm \varepsilon}$ are the photon energy and the polarization vector of the incident light, ${\bf p}$ is the momentum operator, $\Phi$ is the work function. In this work, we focus on nondegenerate molecular orbitals and consider real-valued $\psi_i({\bf r})$. 
When we discuss a 2D PMM for a fixed $E_{k} = h\nu-E_i-\Phi$, the photoelectron intensity is proportional to the squared modulus $I_{\bm k}$ of the transition matrix element defined as 
\begin{eqnarray}
I_{{\bf k}}( \ket{\psi_i}) = |{\bm \varepsilon}\cdot{\bf k}|^2 |\braket{ {\bf k}|\psi_i}|^2.
\label{eq:fermi_tanitu}
\end{eqnarray}

Let us apply Eq. \eqref{eq:fermi_tanitu} to an organic molecular layer. We consider systems consisting of weakly interacting molecules in which the electronic states retain the character of molecular orbitals of an isolated molecule. In such systems, the photoelectrons emitted from individual molecules do not interfere with each other, and the net photoelectron intensity is proportional to the incoherent sum of $I_{{\bf k}}$'s in Eq. \eqref{eq:fermi_tanitu} for different molecules. 
Writing the molecular orbital of the $r$-th molecule as $\ket{\psi_i^r}$ and assuming the binding energy to be the same for all $r$ ($r = 1, 2, \cdots, p$, $\ket{\psi_i^1}=\ket{\psi_i}$), we redefine $I_{\bm k}(\ket{\psi_i})$ as the summation over $r$,
\begin{eqnarray}
I_{{\bf k}}(\ket{\psi_i}) = |{\bm \varepsilon}\cdot{\bf k}|^2 \sum_{r=1}^p  |\braket{{\bf k}|\psi_i^r}|^2. \label{eq:fermi_tochu}
\end{eqnarray}

We can express the right-hand side of Eq. \eqref{eq:fermi_tochu} in terms of a single molecular orbital when the orbitals of different molecules are related to each other by translation and rotation. Setting $\ket{\psi_i^1}=\ket{\psi_i}$ as a reference, $\ket{\psi_i^r}$ is expressed using translation $T_r$ and rotation $R_r$ as follows:
\begin{eqnarray}
\ket{\psi_i^r} = T_r R_r e^{i\theta_r}\ket{\psi_i} \quad (T_1 = R_1 =I).
\label{eq:godo}
\end{eqnarray}
Here $I$ is the identity operator, and $e^{i\theta_{r}}$ is a phase factor. 
Since translations of molecules do not affect the photoelectron intensity, we drop $T_r$ below. 
Substituting Eq. \eqref{eq:godo} into Eq.\eqref{eq:fermi_tochu}, we obtain 
\begin{eqnarray}
I_{{\bf k}}(\ket{\psi_i}) = |{\bm \varepsilon}\cdot{\bf k}|^2 \sum_{R} w_{R} |\bra{ {\bf k}} R \ket{\psi_i}|^2.
\label{eq:fermi_tahaikou}
\end{eqnarray}
Here $R$ indicates different operations, and $w_{R}$ the number of molecules (or the number ratio of molecules) specified by $R$. 
We omit superscript 1 for the reference system.

Finally, we express the photoelectron intensity \eqref{eq:fermi_tahaikou} in terms of the projection operator
\begin{eqnarray}
P =\ket{\psi_i}\bra{\psi_i}, 
\label{eq:P}
\end{eqnarray}
which is convenient for our POT. The intensity is 
\begin{eqnarray}
I_{{\bf k}}(P) = |{\bm \varepsilon}\cdot{\bf k}|^2 {\rm Tr}\left [  F_{{\bf k}} P \right ],
\label{eq:fermi_ope}
\end{eqnarray} 
with 
\begin{eqnarray}
f_{{\bf k}}=\sum_{ R} w_{R} R^{-1}\ket{{\bf k}}\bra{{\bf k}}{R}.
\label{eq:f_ope}
\end{eqnarray}
The projection operator ${P}$ fulfills
\begin{align}
\quad\langle \psi|{P}|\psi\rangle \ge 0, \quad \text{Tr}[P]=1,\quad \text{rank}({P}) = 1,
\label{Eq:P_condition}
\end{align}
where $|\psi\rangle$ is an arbitrary single electron state. 
We denote the first property (positive semidefiniteness) as ${P} \succeq 0$.

To numerically handle Eq. \eqref{eq:fermi_ope}, we approximate $\ket{\psi_i}$ as
\begin{eqnarray}
\ket{\psi_i}=\sum_{n=1}^N \ket{\chi_n} c_n
\label{eq:tenkai}
\end{eqnarray}
with a finite real-valued orthonormal basis $\{{\chi_n}({\bf r})|n=1,...,N\}$.
Regarding the expansion coefficient $c_n$ as the $n$-th component of a normalized real column vector ${\bf c}$, 
\begin{eqnarray}
{{\bf c}}=
\left(
c_{1},
c_{2},
\cdots,
c_{n},
 \cdots,
c_{N}
\right)^T,
\label{eq:c}
\end{eqnarray}
$P$ reduces to a matrix
\begin{eqnarray}
C = {{\bf c}} {{\bf c}}^T.
 \label{Eq:X}
\end{eqnarray}
Thereby, we obtain the matrix representation of Eq. \eqref{eq:fermi_ope} in terms of $C$ as 
\begin{eqnarray}
I_{{\bf k}}(C)&=& |{\bm \varepsilon}\cdot{\bf k}|^2  \sum_{n, n'=1}^N  \bra{\chi_n} f_{{\bf k}} \ket{\chi_{n'}}  c_{n'} c_n 
\nonumber \\
&=& |{\bm \varepsilon}\cdot{\bf k}|^2 \textrm{Tr}\left [F_{{\bf k}} C \right ],
\label{eq:henkei}
\end{eqnarray}
where $F_{{\bf k}}$ in the second line is a matrix whose $n,n'$-component is $\bra{\chi_n} f_{{\bf k}} \ket{\chi_{n'}}$. $C$ inherits the properties of $P$ in Eq. \eqref{Eq:P_condition}: 
\begin{align}
C \succeq 0, \quad \text{Tr}[C]=1,\quad \text{rank}(C) = 1.
 \label{Eq:C_condition}
\end{align}
Eq. \eqref{eq:henkei} with conditions Eq. \eqref{Eq:C_condition} turns out to be a suitable expression for the PhaseLift algorithm as shown below (Sec. \ref{Sec:phaselift}).

\subsection{Photoemission orbital tomography based on least squares fitting}
\label{Sec:LSF}
We aim at determining ${\bf c}$ for $\ket{\psi_i}$ by fitting experimental photoelectron intensities to Eq. \eqref{eq:henkei}.
Suppose that we have experimental photoelectron intensities $z_m$ from an unknown molecular orbital $\ket{\psi_i}$ at ${{\bf k}} = {{\bf k}_m}$ $(m = 1, 2, \cdots, M)$. 
Our problem is a non-linear least squares problem,
\begin{alignat}{2}
&\underset{{\bf c},\lambda}{\text{Minimize}} \quad && \left\Vert  {\bf z}-\lambda{\bf I}({\bf c}{\bf c}^T)\right\Vert_{2}
\label{eq:mondai}
\end{alignat}
where  ${\bf z} = (z_1, z_2, \cdots, z_M)^T$ and 
${\bf I}(A) = (I_{{\bf k}_1}(A), I_{{\bf k}_2}(A) \cdots, I_{{\bf k}_M}(A))^T$. 
$\|\cdot\|_{2}$ represents the Euclidean distance ($L_2$ norm), and
$\lambda$ is the constant of proportionality. 

To solve this problem, we use the trust region reflective (TRF) algorithm \cite{TRF} and the basin hopping algorithm \cite{Basin}.
In the least squares fitting, the results can be at local minima depending on the initial values of $\bf c$.

\subsection{Orbital tomography based on PhaseLift}
\label{Sec:phaselift}
For POT, we can also use a PhaseLift based algorithm.
PhaseLift algorithm is classified as a semidefinite program (SDP) and is known to be robust against additive noise \cite{CSV13}.

In the PhaseLift based approach, we find a low-rank real symmetric matrix $\overline C$ instead of $\lambda$ and $\bf c$ by minimizing $\text{Tr}[{\overline C}]$ under the experimental constraints.
We solve the following problem with a finite uncertainty $\eta$,
\begin{alignat}{2}
 &\underset{\overline{C}}{\text{Minimize}} \quad && {\text{Tr}[\overline{C}]}    \nonumber  \\
 &\text{subject to}             \quad && \left\Vert {{\bf z}} - {{\bf I}}(\overline{C}) \right\Vert_2 \le \eta, \nonumber\\
 &{}                            \quad && \overline{C} \succeq 0.
\label{eq:CSV13}
\end{alignat}
In the expression $I(\overline C)$, the constant of proportionality $\lambda$ is included in the matrix $\overline C$. 
In reality, there can  be contributions to $\eta$ other than
the statistical noise, such as systematic uncertainties in the data due to, for example,
signals from the substrate, or radiation damage of the sample.
There are also uncertainties which arise from the theoretical model
such as use of an insufficient basis function set, errors in the assumed molecular structure, or
the plane wave approximation of the photoelectron final states.

The rank of resulting $\overline{C}$ can be larger than 1. In that case, $\overline{C}$ represents a mixed state containing more than one component. 
Following Ref. \cite{CSV13}, we extract the rank 1 matrix $C$ by eigenvalue decomposition
\begin{eqnarray}
 \overline{C} &=& \sum_{i=1}^{N} \nu_{i} {\bf u}_{i} {\bf u}_{i}^{T},\quad \nu_1 \ge \nu_2 \ge \cdots \ge \nu_N \ge 0.
 \label{eq:EVD}
\end{eqnarray}
Here $\nu_{i}$ and ${\bf u}_{i}$ are the eigenvalues and normalized eigenvectors of the matrix $\overline{C}$. We regard ${\bf u}_1$ as $\bf c$, 
\begin{eqnarray}
C={\bf u}_1{\bf u}_1^T.
 \label{Eq:C_phaselift}
\end{eqnarray}
In order to correct for the loss of the intensity caused by the disregard of the minor components, we rescale the intensity by minimizing the distance between $\bf z$ and rescaled ${\bf I}(C)$, $\left\Vert {\bf z} - \lambda {\bf I}(C)\right\Vert_{2}$.
The distance is minimum for $\lambda = \frac{{\bf z}\cdot {\bf I}(C)}{\left \|{\bf I}(C)\right \|_2^2}$ where it has the value
\begin{eqnarray}
d(C)={ \left \|{\bf z}\right \|_2 }
 \left ( 1 - \frac{\left({\bf z}\cdot {\bf I}(C)\right)^2}{ { \left \|{\bf z}\right \|_2^2}  \left \|{\bf I}(C)\right \|_2^2 } \right )^{1/2}.
\label{eq:d}
\end{eqnarray}
Replacement of $\nu_1$ by the optimized $\lambda$ is similar to the ``debiasing'' of the solution in Ref. \cite{CSV13}.  This is also useful when $\overline{C}$ is rank 1 because intensities tend to be scaled down through trace minimization, especially at large $\eta$. 
Since the appropriate value of $\eta$ for each particular case is not known a priori, as a practical means for choosing $\eta$, we find $C$ with the smallest $d(C)$ by repeating the same procedure with different $\eta$'s.

\subsection{Atomic orbital basis set}
Since the molecular orbitals of our interest are localized around the molecule, they can be approximated with a small number of atomic orbitals \cite{Szabo}. 
Atomic orbitals $\{\ket{\phi_{n}}|n = 1, 2, \cdots, N \}$ are usually nonorthogonal, yet we can easily orthonormalize them. 
To this end, we can use, for example, L\"owdin's linear transformation \cite{LP50, Szabo}.
Diagonalizing the overlap matrix $S_{nn'} = \langle \phi_n| \phi_{n'}\rangle$,
\begin{align}
U^T SU &= s, 
\end{align}
we define matrix V as 
\begin{align}
 V &= U s^{-\frac{1}{2}}.
\end{align}
$U$ is an orthogonal matrix composed of the eigenvectors of $S$,
$s$ the diagonal matrix whose diagonal elements are the eigenvalues of $S$, 
and $s^{-\frac{1}{2}}$ is the diagonal matrix of the inverse square root of the eigenvalues of $S$. 
With $V$, we can orthonormalize the atomic orbitals as
\begin{eqnarray}
\ket{\phi_{n}'} = \sum_{q} \ket{\phi_q} V_{q n}.
\label{eq:LO} 
\end{eqnarray}
We use these orthonormal orbitals $\{ \ket{\phi_n'}\}$ as $\{\ket{\chi_n}\}$ in Eq. \eqref{eq:tenkai}.

\section{Application}
\label{sec:data}
\begin{figure}[tb]
\begin{tabular}{ll}
(a)&(b)\\
 \includegraphics[width=4.8cm]{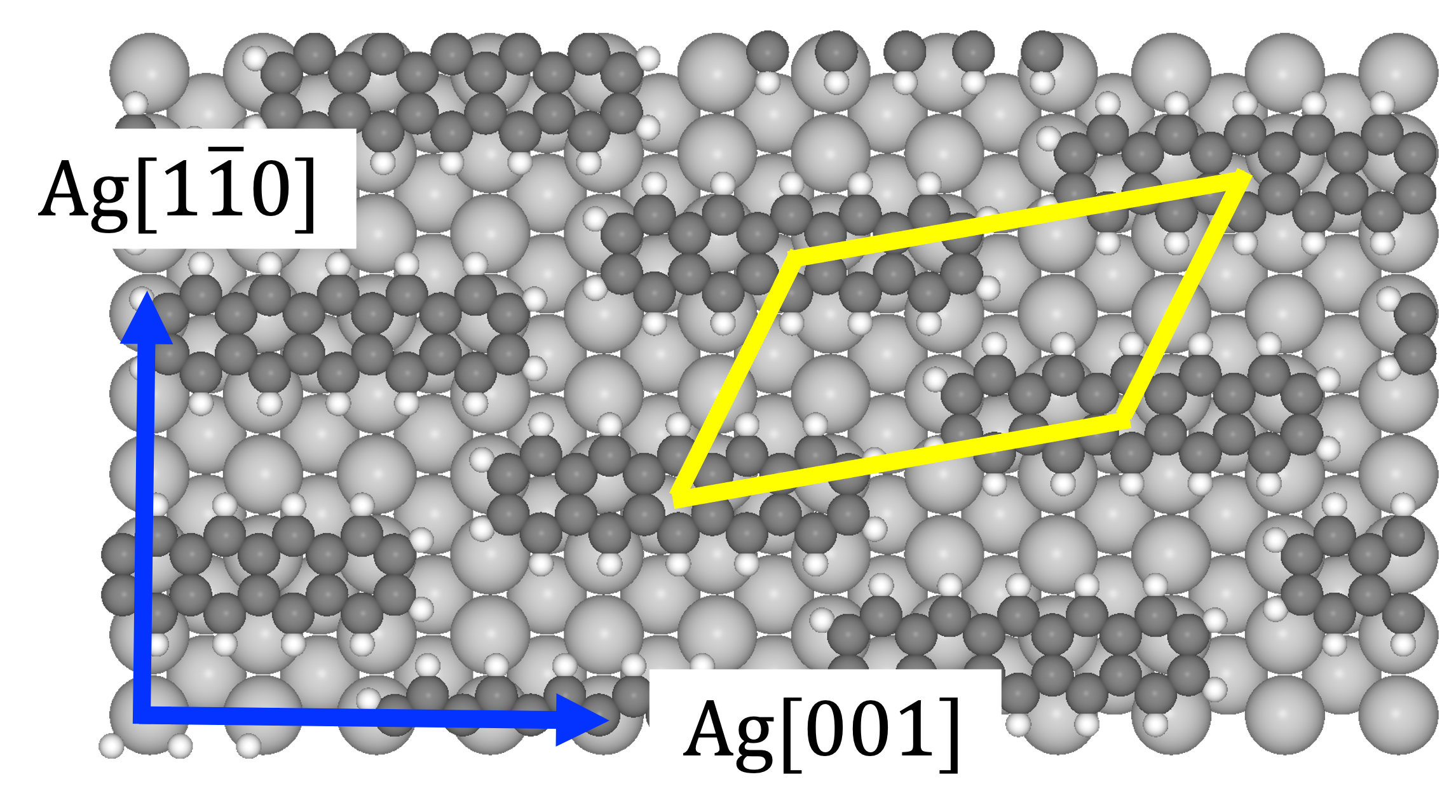}
 & \includegraphics[width=3.7cm]{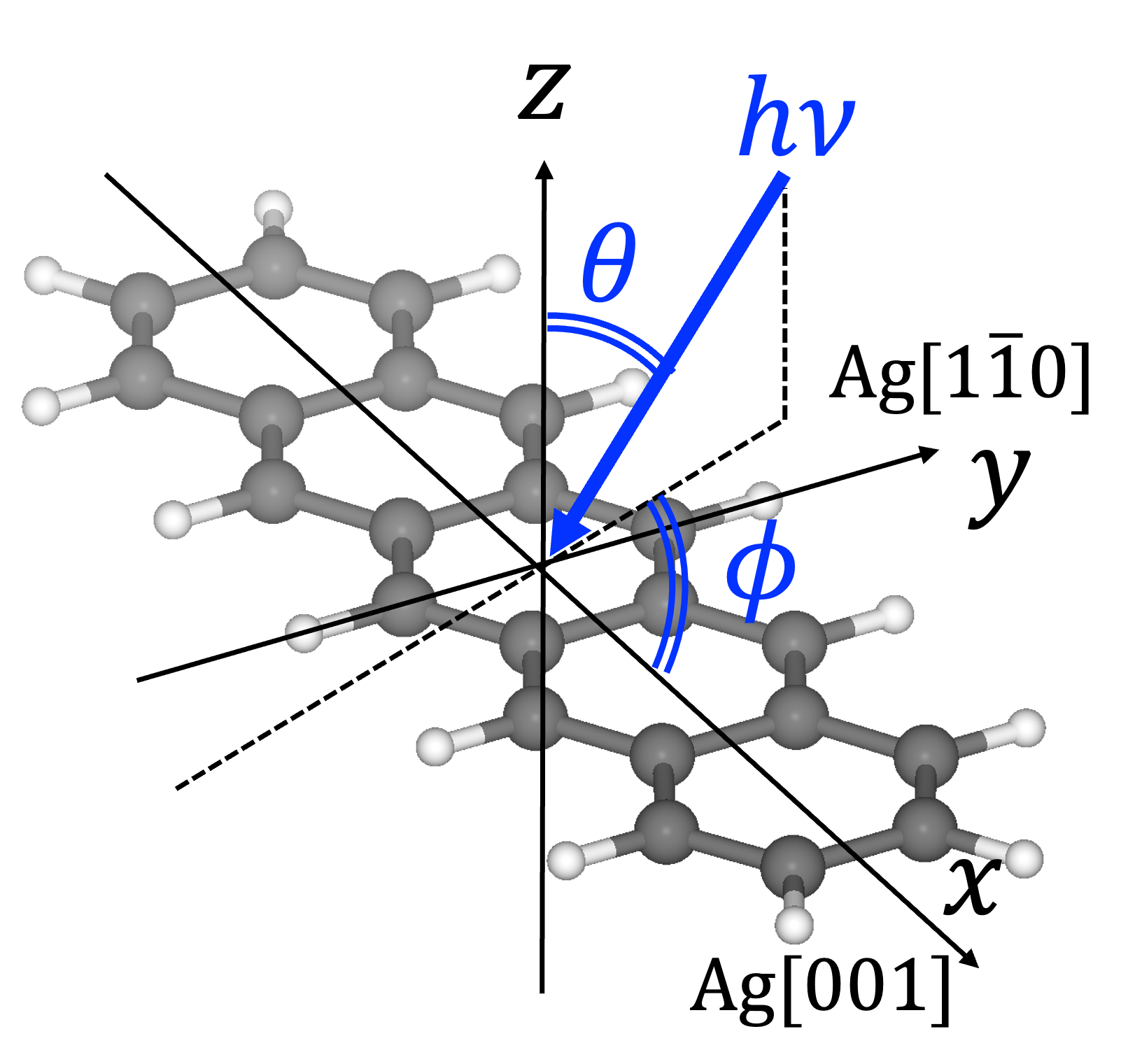}\\
(c)&(d)\\
 \includegraphics[width=4.8cm]{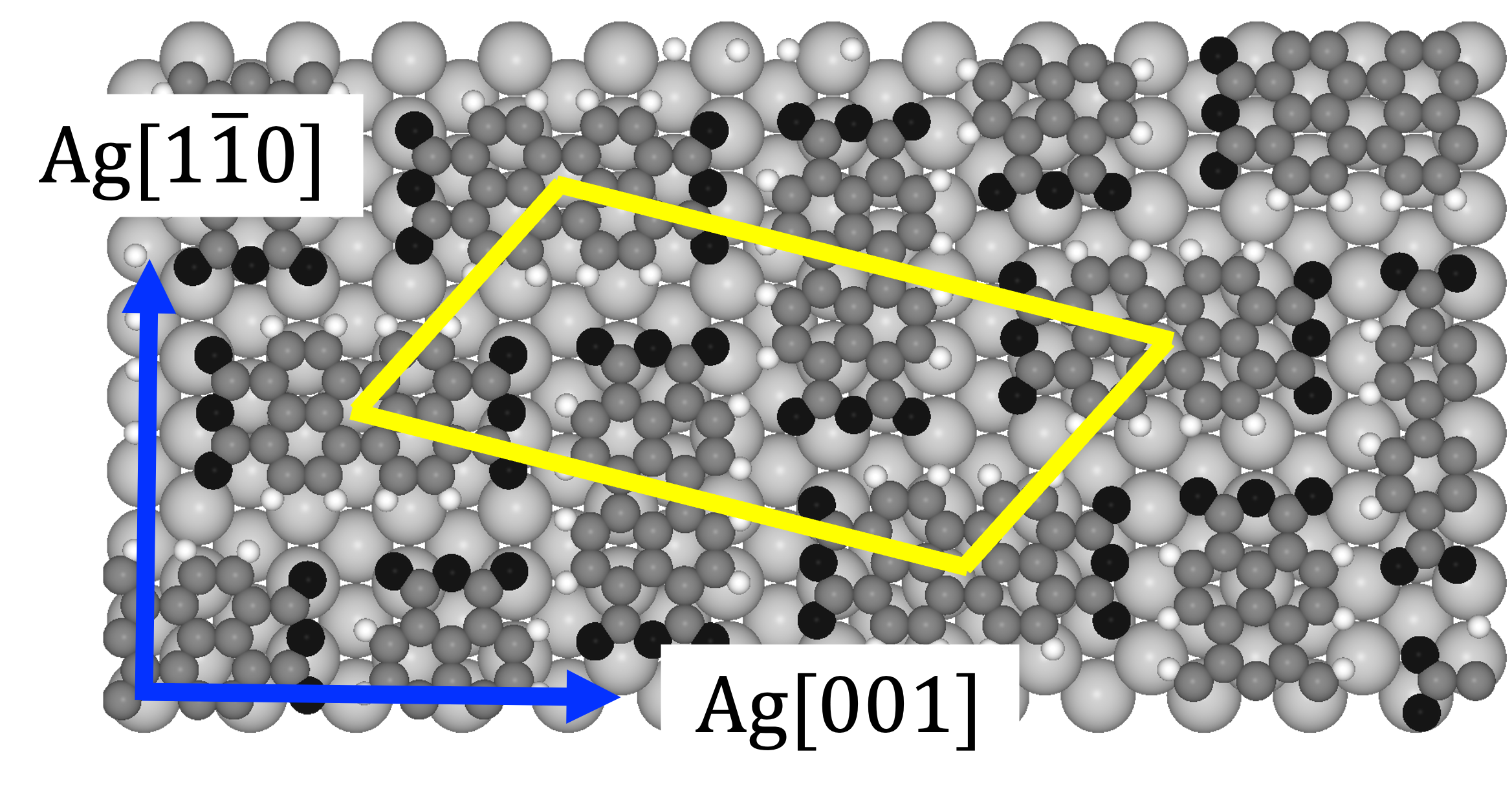}
 & \includegraphics[width=3.7cm]{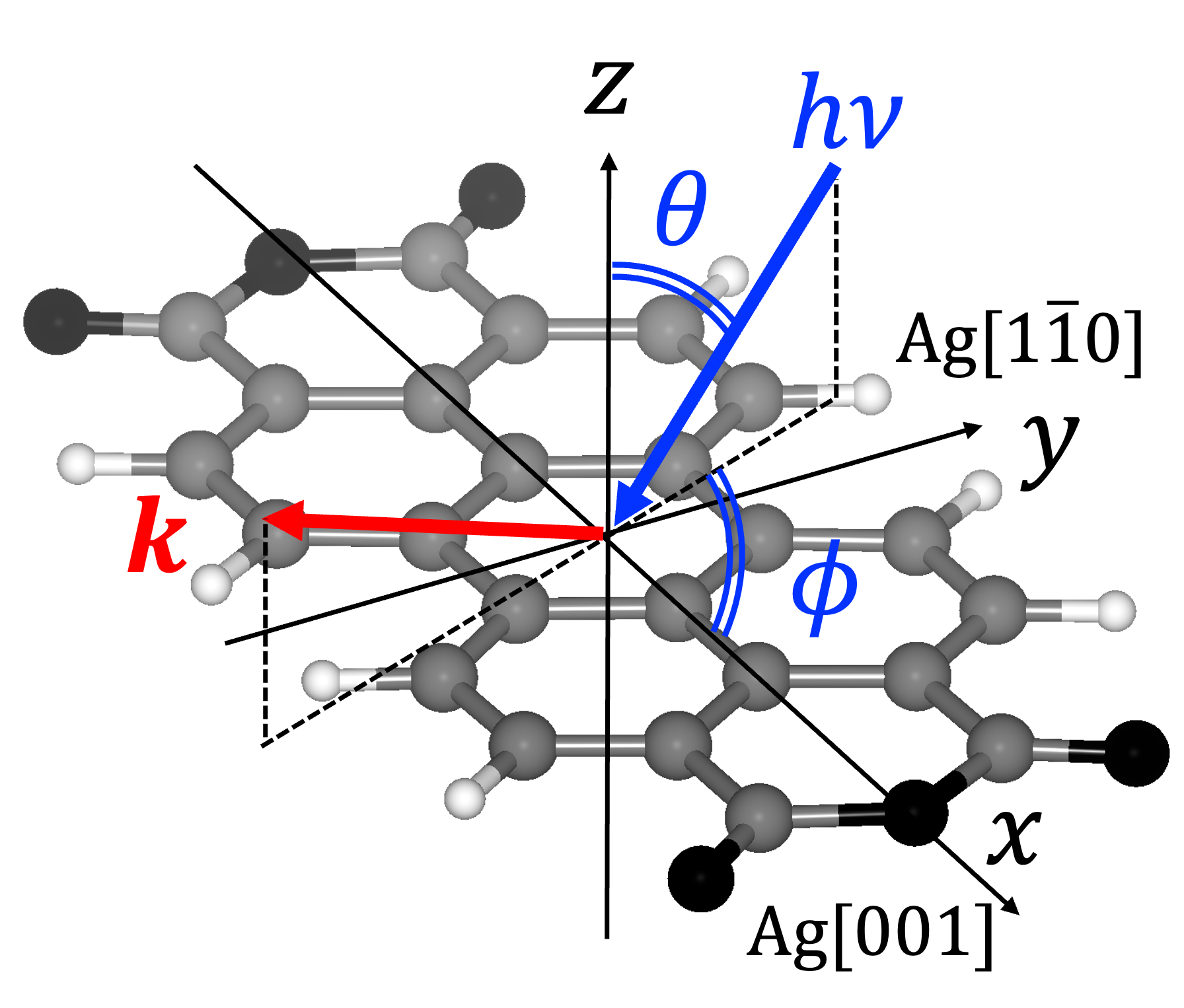}
 \end{tabular}
 \caption{(a),(c) Structural models of (a) pentacene/Ag(110) and (c) PTCDA/Ag(110). The dark gray, black, white, and light gray spheres are, respectively, carbon, oxygen, hydrogen, and silver atoms. 
 The parallelograms in (a) and (c) represent surface unit cells. 
 (b),(d) The relation between the molecular coordinates and the incident direction of the light for (b) pentacene/Ag(110) and (d) PTCDA/Ag(110) in the measurement in Ref. \cite{Metzger2021} and Ref. \cite{WSS13}.
 For PTCDA/Ag(110) system, the incident direction of the light and photoelectron wave vector ${\bf k}$ reside in the same plane perpendicular to the molecular plane.}
\label{fig:geometry}
\end{figure}

In this work, we apply the above method to the highest occupied molecular orbital (HOMO) of monolayer pentacene on Ag(110) (single-orientation system) and the HOMO of herringbone monolayer PTCDA on Ag(110) surfaces (multi-orientation system). 
Figures \ref{fig:geometry}(a) and \ref{fig:geometry}(c) illustrate surface models of monolayer pentacene/Ag(110) and monolayer PTCDA/Ag(110), respectively. All the pentacene molecules lie flat on the silver surface and are aligned with their long molecular axis along the [001] direction of the silver substrate \cite{WJS04, ULR14}. 
All the PTCDA molecules are lying flat on the silver surface, half of them aligned with their long molecular axis along the [001] and the other half along the [1$\bar 1$0] direction of the silver substrate \cite{WSS13, AS09}. To our knowledge, their precise adsorption positions have not been determined experimentally. 
In both systems, the HOMO does not show significant hybridization with the substrate \cite{WZF13, ZFS10, AS09, ULR14}.

\subsection{Experimental data}
\subsubsection{pentacene/Ag(110)}
\label{Sec:PMM_pen}
As the experimental PMM of the HOMO of pentacene/Ag(110), we take the experimental data set by Metzger {\it et al.} reported in Ref. \cite{Metzger2021}. 
In their experiment, photoelectrons excited from the initial state at the binding energy $E_i$ = 1.2 eV with the circularly polarized light of $h\nu$ = 28 eV incident from the polar angle $\theta$ = 65$^\circ$ and azimuthal angle $\phi$ = -85$^\circ$ are collected [Fig. \ref{fig:geometry}(b)]. 
As the value of the work function is not reported in Ref. \cite{Metzger2021}, 
we use the value $\Phi$ = 4.0 eV and $E_k$ = 22.8 eV. 
This arbitrary choice of $E_k$ does not have a major effect on the orbital reconstruction except for the quantitative molecular structure 
determination described in Sec.\ref{Sec:v0}. See Appendix \ref{Sec:appeMO1} for further information. 
The data which we use here [Fig. 4.5(b) in Ref. \cite{Metzger2021}] have been background-substracted and symmetrized with respect to $k_x$ and $k_y$ axes. 
According to Ref. \cite{Metzger2021}, they extracted the signals from the HOMO of pentacene from the raw data by fitting the energy distribution curve at each data point assuming a Gaussian line shape for each molecular orbital and an exponential function plus constant for the substrate
signals, multiplied by the Fermi function.

We prepared the input experimental PMM data for our analysis shown in Fig. \ref{fig:pen}(a) from the data in Ref. \cite{Metzger2021} in two steps. 
In the first step, we extracted numerical values from the 2D color map of PMM [Fig. 4.5(b) in Ref. \cite{Metzger2021}]. 
In the second step, we performed an average pooling over the range of $\Delta k_x \Delta k_y$ $\approx$ 0.07$^2$ \AA$^{-2}$ on the digitized experimental PMM.

In the estimation of the molecular orbital, it is necessary to take into account the experimental condition and the effect of the above-mentioned data processing. The symmetrization of the data is equivalent to the superposition of the four PMMs with incident light from $(\theta, \phi) = (65^\circ, -85^\circ), (65^\circ, 85^\circ), (65^\circ, 95^\circ)$, and $(65^\circ -95^\circ)$.
We take the sum of $|{\bm \varepsilon} \cdot {\bf k}|^2$ for the four incident directions in Eq. \eqref{eq:henkei}.

\subsubsection{PTCDA/Ag(110)}
As the experimental PMM for the HOMO of PTCDA/Ag(110), we refer to the experimental data 
by Willenbockel {\it et al.} reported in Ref. \cite{WSS13}. 

In their experiment, the photoelectrons are excited from the initial state at $E_i$ = 1.74 eV by the $p$-polarized light of $h\nu$ = 35 eV with incident angle $\theta$ = 40$^\circ$. The sample was rotated around the $z$-axis, and the photoelectrons emitted in the plane spanned by the $z$-axis and the direction of the incident light were detected [Fig. \ref{fig:geometry}(d)]. 
As the value of $\Phi$ is not reported in Ref. \cite{WSS13}, we use the value $\Phi$ = 4.0 eV and $E_k$ = 29.3 eV. There is no description about the background subtraction in Ref. \cite{WSS13} and the data may contain contributions from other than PTCDA molecules. 

We digitized the PMM in the same manner as pentacene/Ag(110) described in Sec. \ref{Sec:PMM_pen}. 
In our orbital estimations, we assume that the orientations of the two molecules in the surface unit cell are related to each other by 90$^\circ$ rotation. 
According to the analysis in Ref. \cite{WSS13}, contributions from the rotated molecules to the PMM at $E_i$ = 1.74 eV are not 1:1 due to a small difference in the binding energies of the rotated molecules and the presence of a small amount of another phase. 
In order to reduce the number of fitting parameters, we symmetrized the PMM with respect to the line $k_x=k_y$ and fix the contributions of the rotated molecules to be 1:1 in the subsequent analysis. 
The symmetrized PMM data is shown in Fig. \ref{fig:pt}(a).  

\subsection{Computational details}

\subsubsection{Density functional theory calculation}
We performed structural optimization for isolated pentacene and PTCDA at the level of B3LYP \cite{Becke1993} with the 6-31++G(d,p) basis set using Gaussian16 \cite{g16}. The obtained structures were used in the estimations of molecular orbitals (Sec. \ref{Sec:pen} and Sec. \ref{Sec:pt}), and the Kohn-Sham (KS) orbitals were compared with the estimated orbitals [Fig. \ref{fig:pen}(e) and Fig. \ref{fig:pt}(d)].

For comparison with the optimized structure using PMM, we also performed the structural optimization for pentacene monolayers adsorbed on Ag(110) surface using the Vienna Ab initio Simulation Package (VASP) \cite{vasp}. We used the projector-augmented wave (PAW) \cite{paw} approach with an energy cutoff of 400 eV. 
The PBE functional \cite{pbe} was chosen for exchange-correlation effects, and van der Waals corrections are treated with the DFT-D3 method of Grimme with zero-damping function \cite{dftd3zero}. 

\begin{figure}[tb]
\begin{tabular}{ll}
 (a) & (b)\\
\includegraphics[width=4.2cm]{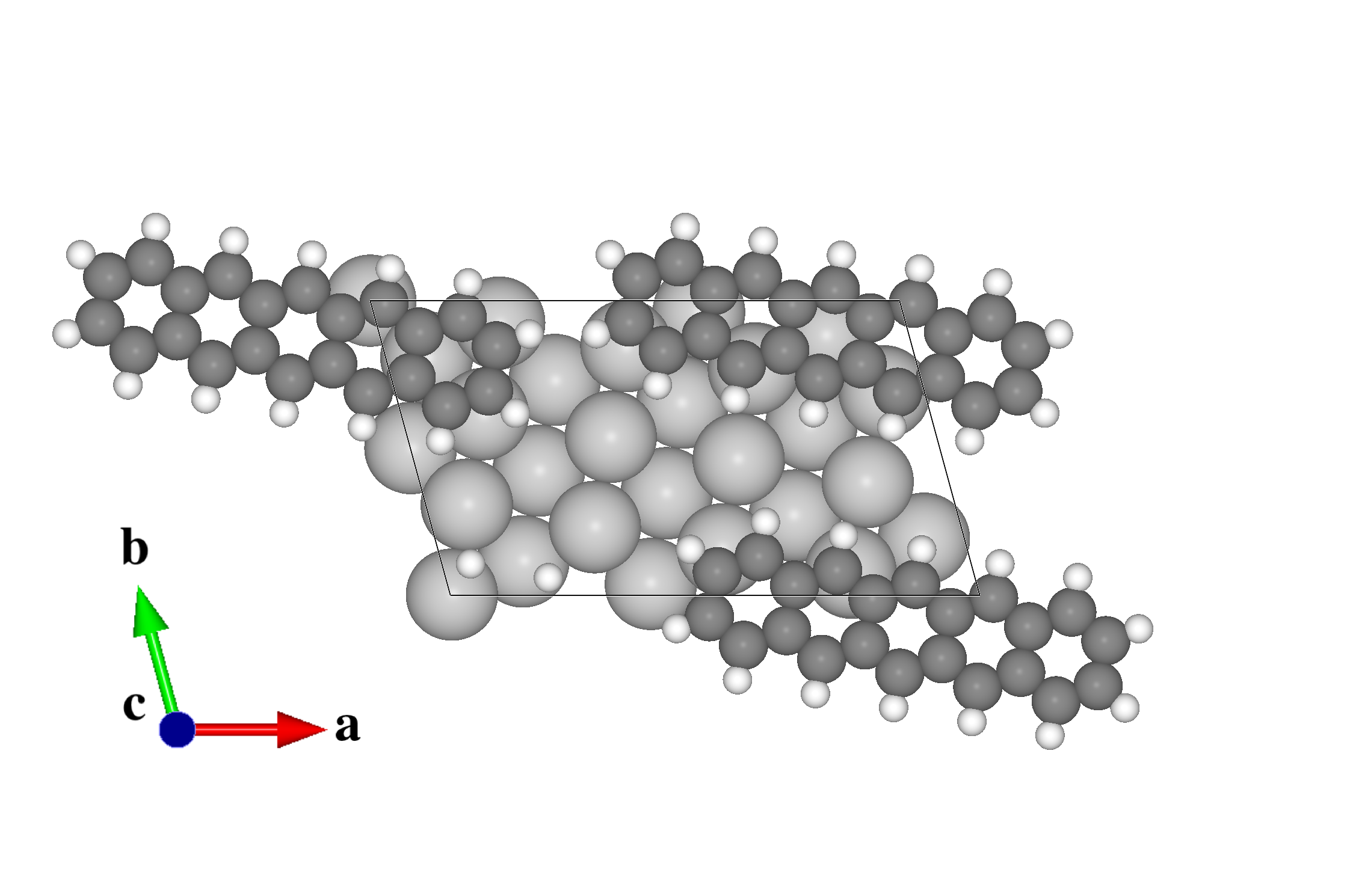}
&
\includegraphics[width=4.2cm]{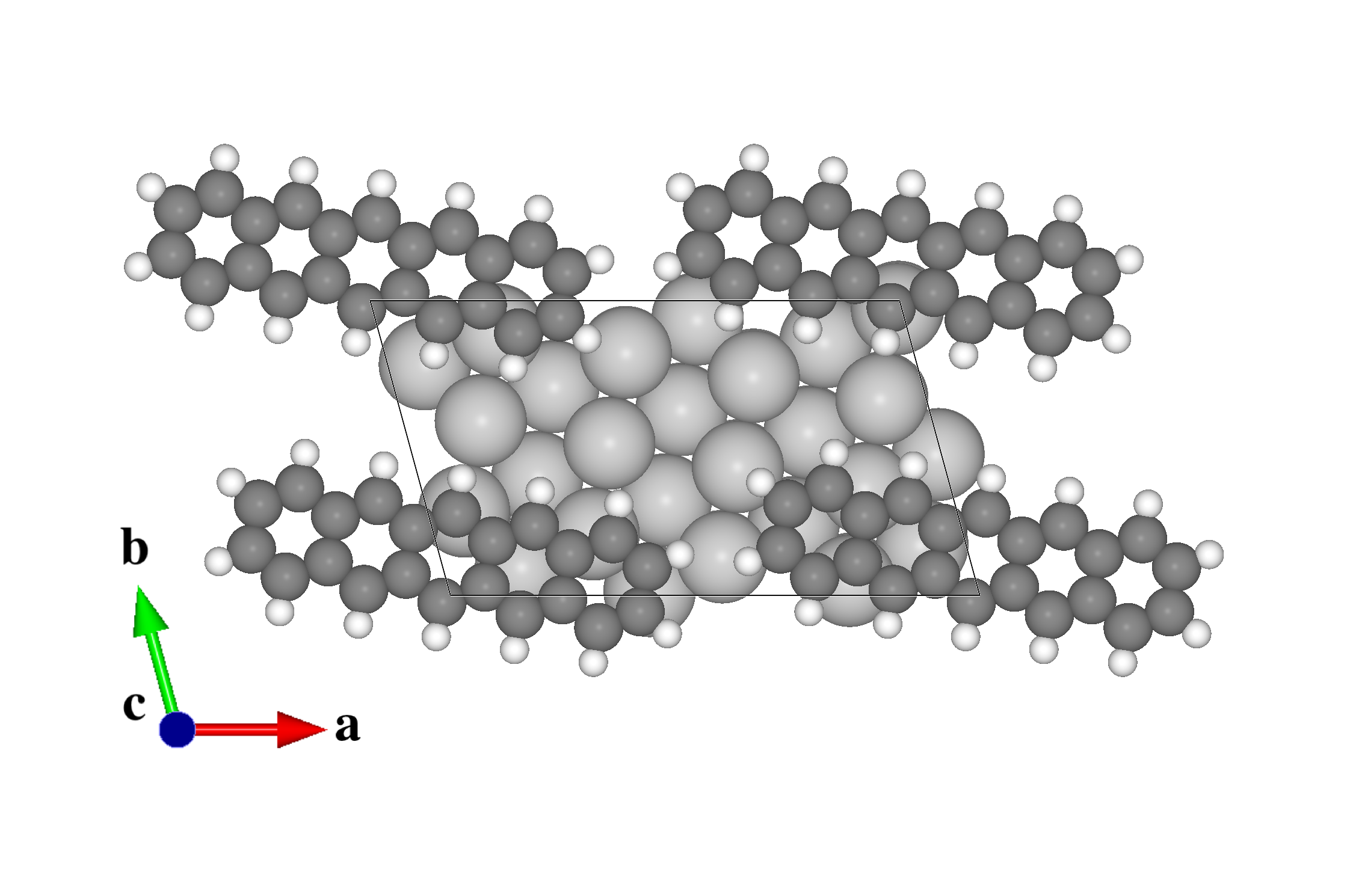}
\end{tabular}
\caption{Model surface structures of pentacene/Ag(110). In (a) and (b), the centroid of pentacene is at the hollow site and short-bridge site, respectively.}
\label{fig:adsorption}
\end{figure}

We performed the calculations for two
possible adsorption positions [Fig. \ref{fig:adsorption}]. 
The surface structure was modeled using a 10-layer Ag(110) slab with a pentacene molecule separated by a vacuum layer of 15 {\AA}. The 2D lattice vectors are described using the superstructure matrix $\begin{pmatrix} 3 &-1\\1 &4\end{pmatrix}$ \cite{ULR14} with the substrate 2D lattice vectors (a,0) and (0,a/$\sqrt{2}$) (a = 4.07 \AA).

In the electronic structure calculations, we used a first-order Methfessel-Paxton smearing \cite{smearing} of 0.2 eV, 3$\times$3$\times$1 Monkhorst-Pack k-mesh grid \cite{monk}, and the energy convergence tolerance of 10$^{-7}$ eV/unit cell. We considered the electronic dipole correction along the c-axis to take into account the possible effect of charge interactions between the repeated slabs.  In the structural optimizations, all the ionic positions were optimized until all residual forces were less than 10$^{-2}$ eV/\AA.     

\subsubsection{POT}
As the input molecular structures, we used the optimized structures of isolated pentacene and PTCDA within the density functional theory (DFT). As the atomic orbitals $\phi_{q}$ in  Eq. \eqref{eq:LO}, we used the STO-3G basis set and considered only 2$p_z$ orbitals of carbon and oxygen atoms in the molecules, omitting the other basis functions in the set. The ground for using the minimal basis set is given in Appendix \ref{Sec:appe}. 
In the orbital estimation of PTCDA with herringbone structure, $R$ in Eq. \eqref{eq:f_ope} includes the identity operator and the 90$^\circ$ rotation operator around the z-axis. $w_R$ for both operators are fixed to be 1. In the least squares fitting, results can fall into local minima depending on the initial values. 
We checked that the TRF and the basin hopping algorithms give the same result by adequately choosing the initial values. In the PhaseLift based approach, we varied $\eta'$ ($\eta=\eta'\cdot\|{\bf z}\|_2$) from 0 to 0.95 by 0.05 step and solved problem \eqref{eq:CSV13} using Splitting Conic Solver (version: 3.1.0) \cite{scs} via cvxpy (version: 1.1.17)  \cite{cvxpy1, cvxpy2}. 

In the orbital reconstruction, 
we imposed the constraints that the molecular orbital coefficients 
are symmetric or anti-symmetric with respect to the reflection through the $yz$ and $xz$ plane. 
These constraints, which are required by the structural symmetry of both pentacene and PTCDA, 
were necessary for the orbital reconstruction from the present data set.

\section{Results and Discussion}
\label{results}
\subsection{Orbital estimation}
\label{Sec:penpt}
\subsubsection{Pentacene/Ag(110)}
\label{Sec:pen}
\begin{figure*}[tb]
\begin{tabular}{llll}
(a) &(b)&(c)&(d)\\
\includegraphics[width=0.2\linewidth]{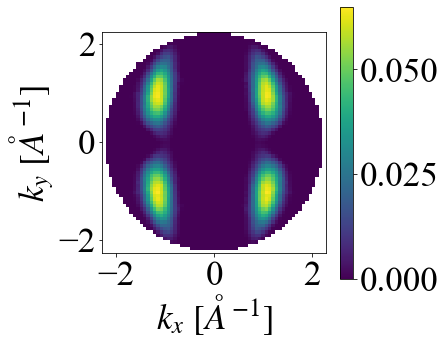}&
\includegraphics[width=0.2\linewidth]{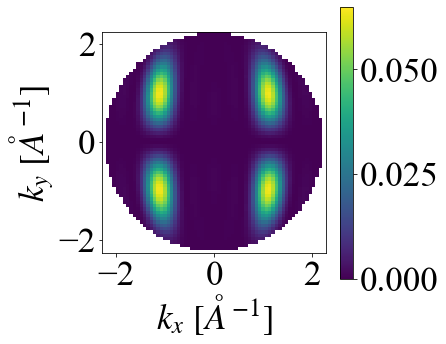}&
\includegraphics[width=0.2\linewidth]{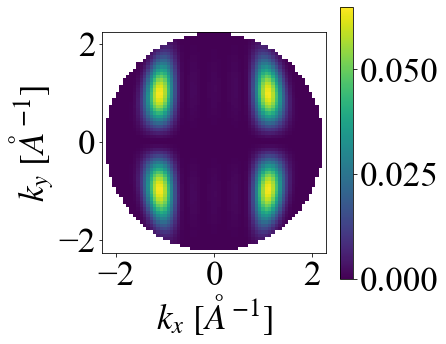}&
\includegraphics[width=0.2\linewidth]{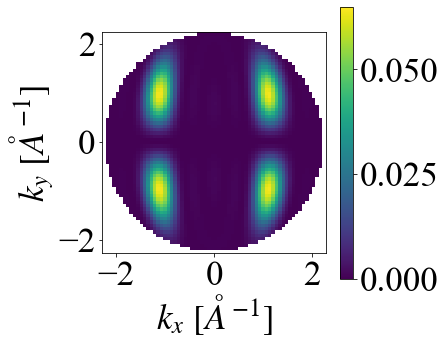}\\
(e)&(f)&(g)&(h)\\
\includegraphics[width=0.2\linewidth]{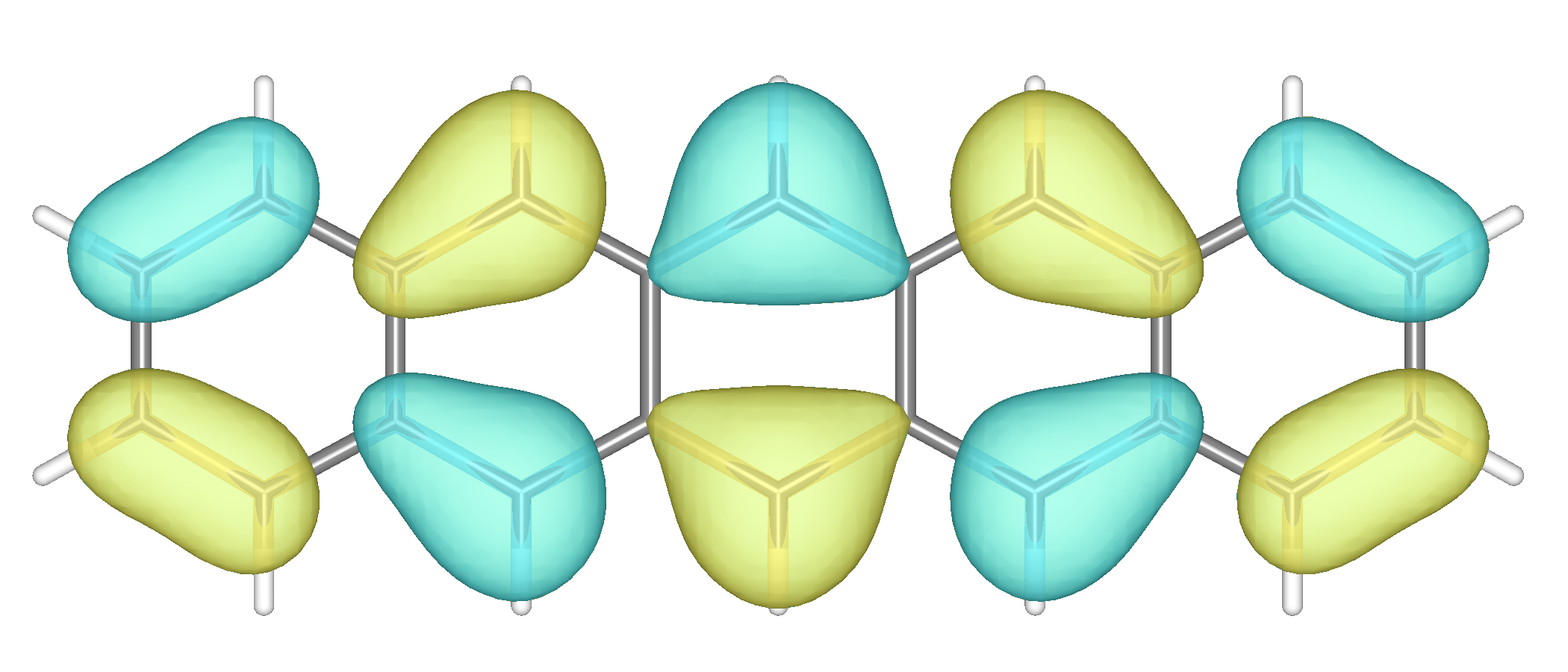}&
\includegraphics[width=0.2\linewidth]{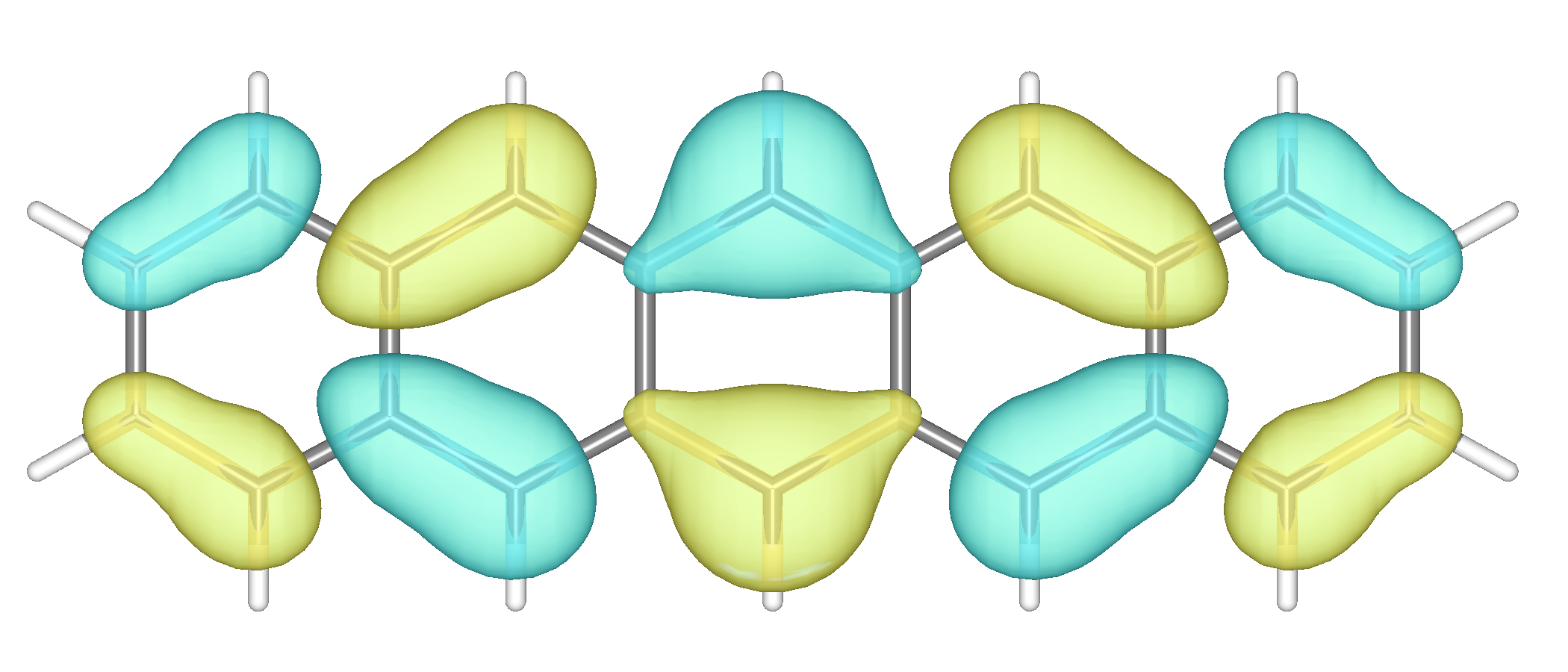}&
\includegraphics[width=0.2\linewidth]{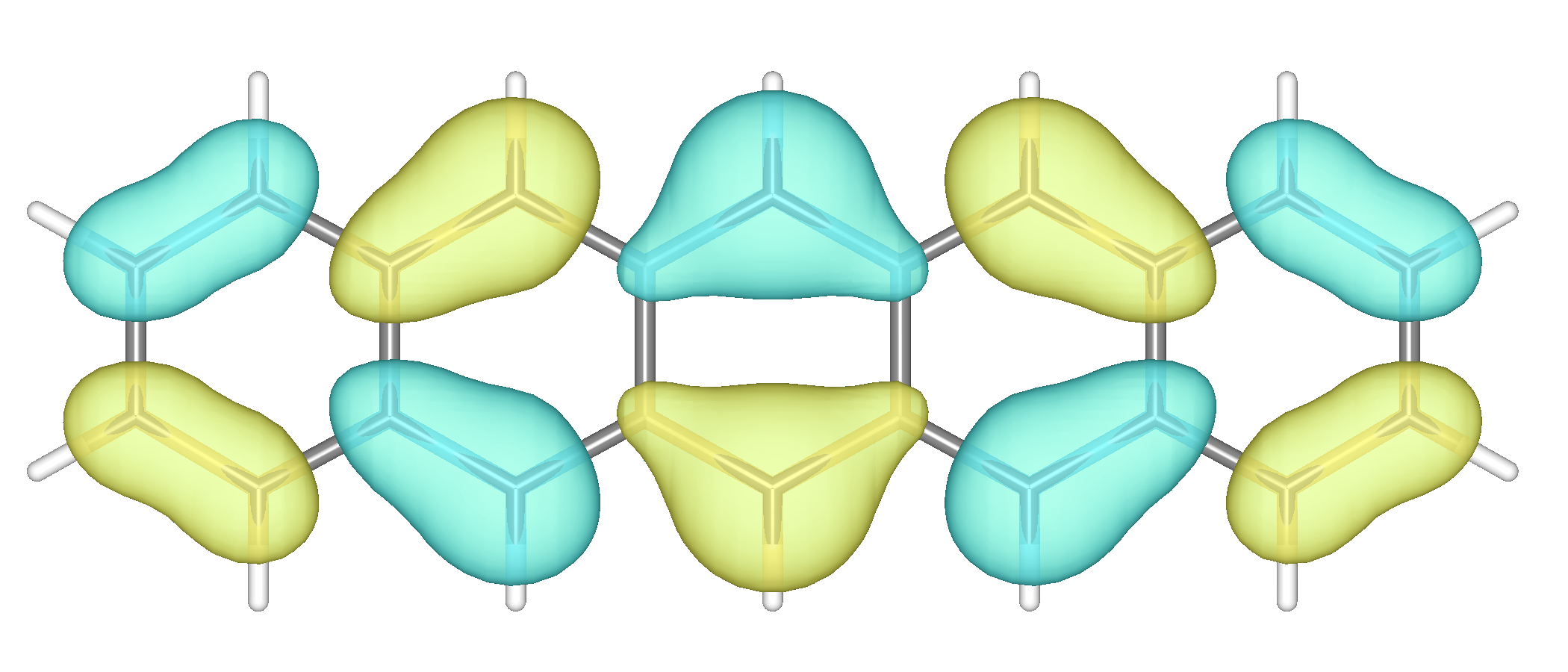}&
\includegraphics[width=0.2\linewidth]{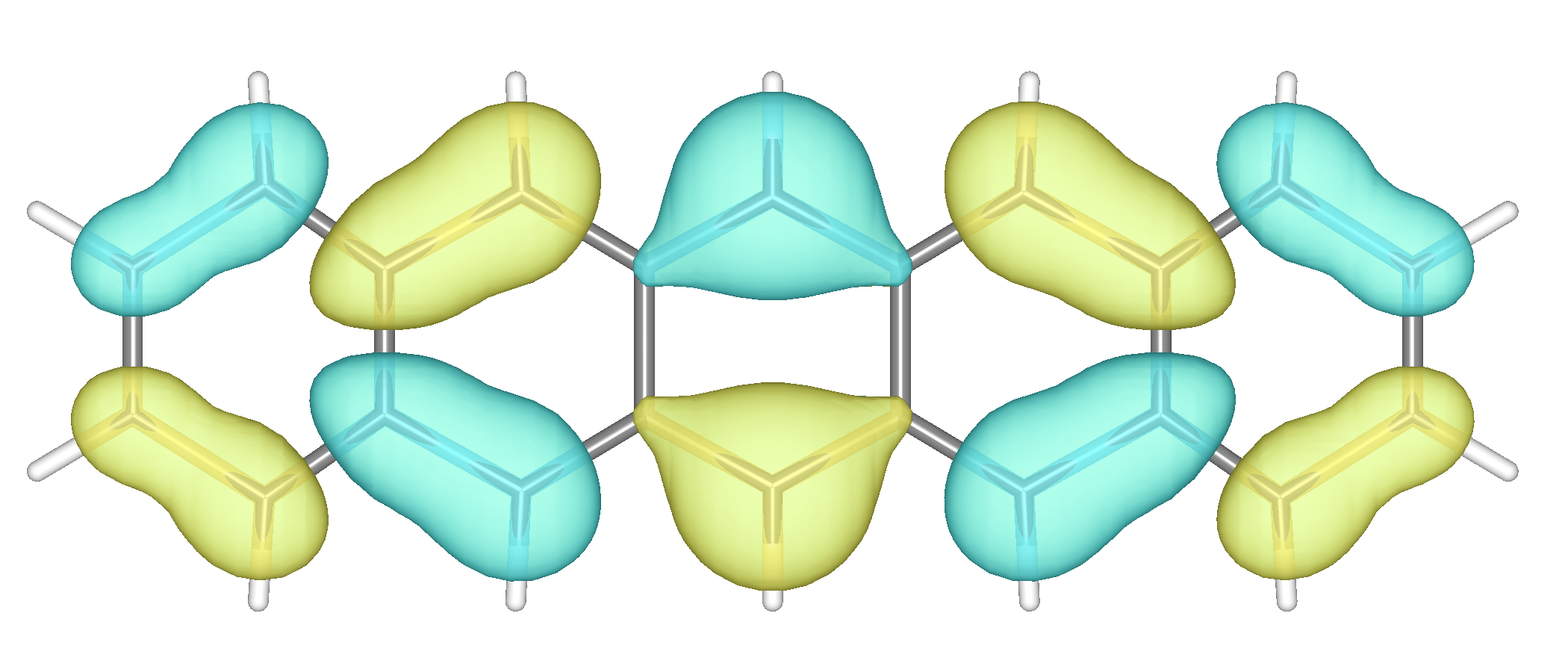}
\end{tabular}
\caption{
(a) Experimental PMM of the HOMO of pentacene/Ag(110) (original data were taken from Fig. 4.5(b) in Ref. \cite{Metzger2021}). 
(b) The fitted PMM with the least squares approach.
(c) The fitted PMM $\lambda {\bf I}(C)$ with the PhaseLift approach at $\eta'$ = 0.20.
In (b) and (c), the DFT-optimized molecular structure was used.
(d) The fitted PMM with the least squares approach for the PMM-optimized structure. 
The plotted data are normalized as ${\bf z}/\|{\bf z}\|_{2}$ for (a) and $\lambda {\bf I}(C)/\|{\bf z}\|_{2}$ for (b), (c) and (d).
(e) The KS-HOMO of an isolated pentacene. 
(f) The estimated orbital with the least squares approach.
(g) The estimated orbital with the PhaseLift based approach at $\eta'$ = 0.20.
(h) The estimated orbital with the least squares approach for the molecular structure optimized using PMM.
In (e)-(h), the isosurface value is 0.02 a.u.
The orbitals are visualized using the VESTA software \cite{VESTA}.  }
\label{fig:pen}
\end{figure*}

We estimated the HOMO of pentacene by using the least squares approach (Sec. \ref{Sec:LSF}).
Figures \ref{fig:pen}(b) and \ref{fig:pen}(f) show the fitted PMM and the estimated orbital. We could well reproduce the experimental PMM with the estimated orbital. The normalized distance between the experimental PMM [Fig. \ref{fig:pen}(a)] and the fitted PMM $\|{\bf z} - \lambda {\bf I}({\bf c}{\bf c}^T)\|_{2}/\|{\bf z}\|_2$ is 0.171. The spatial pattern of the estimated orbital and the KS-HOMO [Fig. \ref{fig:pen}(e)] are consistent with each other.

\begin{figure}[tb]
 \includegraphics[width=6cm]{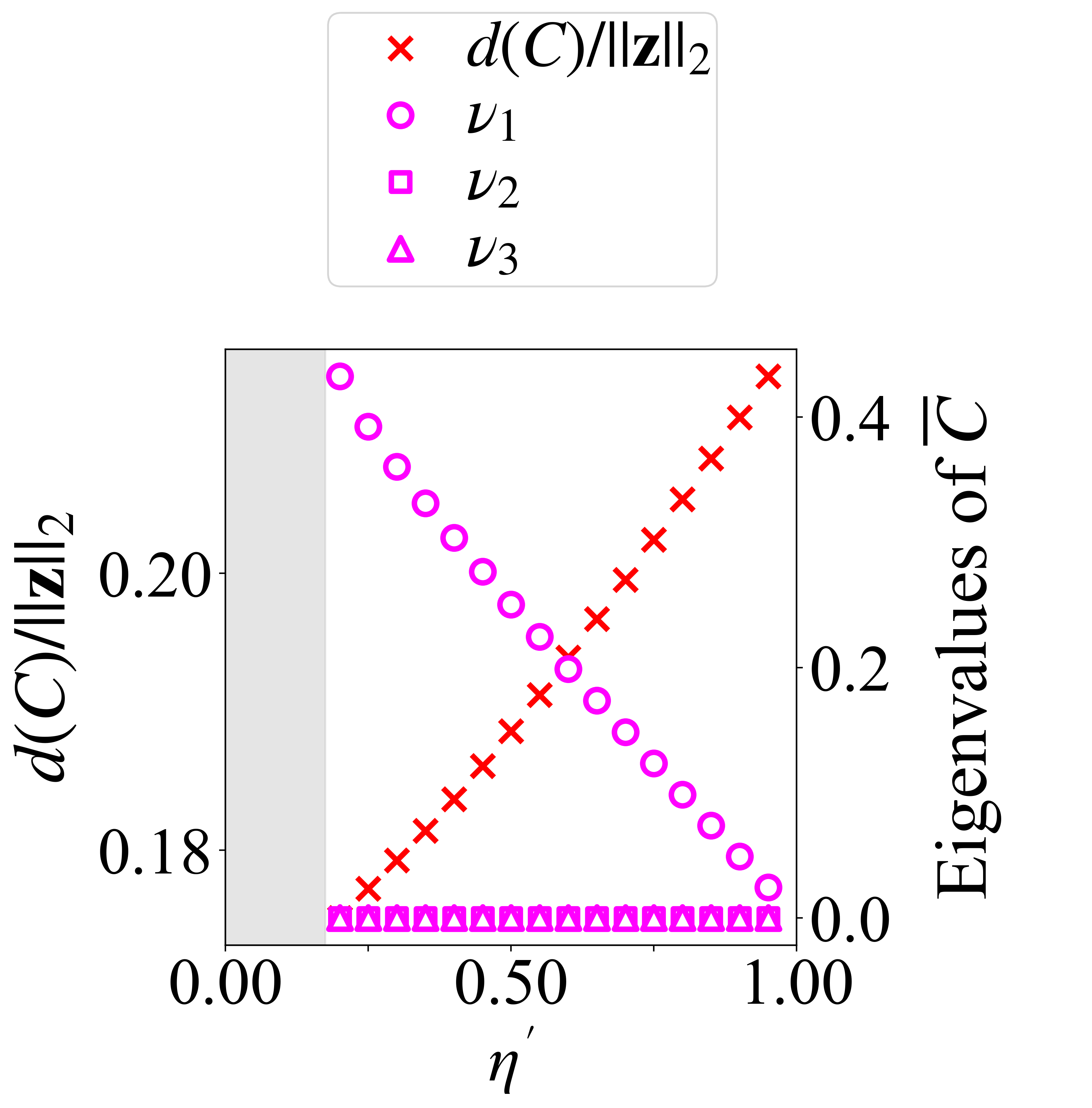}
 \caption{$d(C)$ and eigenvalues of $\overline{C}$ for pentacene/Ag(110). Crosses show the distance between the experimental PMM and fitted PMM ($d(C)$, left axis). Open symbols show the largest three eigenvalues of $\overline{C}$ (right axis).
 In the shaded region, we could not obtain $\overline{C}$.}
\label{fig:pen2}
\end{figure}

We also estimated the HOMO of pentacene by using the PhaseLift based approach (Sec. \ref{Sec:phaselift}). $d(C)/\|{\bf z}\|_2$ takes the minimum value 0.175 at $\eta'$ = 0.20 (Fig. \ref{fig:pen2}). Figures \ref{fig:pen}(c) and \ref{fig:pen}(g) show the fitted PMM $\lambda{\bf I}(C)$ and the estimated orbital at $\eta'$ = 0.20. We could well reproduce the experimental PMM with the estimated orbital [Fig. \ref{fig:pen}(g)]. The estimated orbital resembles KS-HOMO [Fig. \ref{fig:pen} (e)] as well as the estimated orbital of the least squares approach.

\subsubsection{PTCDA/Ag(110)}
\label{Sec:pt}
In the previous application on pentacene, we have shown that with our method we can obtain the three-dimensional molecular orbital in a singe-orientational system, with an accuracy comparable to the Fourier-based POT. In this section, we apply our method to a multi-orientational system, namely PTCDA on Ag(110), where the Fourier-based POT cannot be applied.

\begin{figure*}[tb]
\begin{tabular}{lll}
(a) &(b)&(c)\\
\includegraphics[width=0.25\linewidth]{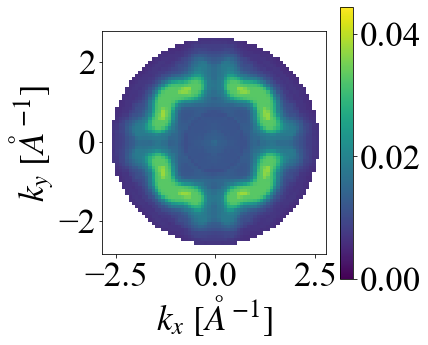}&
\includegraphics[width=0.25\linewidth]{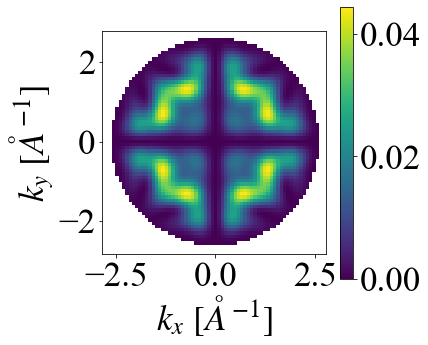}&
\includegraphics[width=0.25\linewidth]{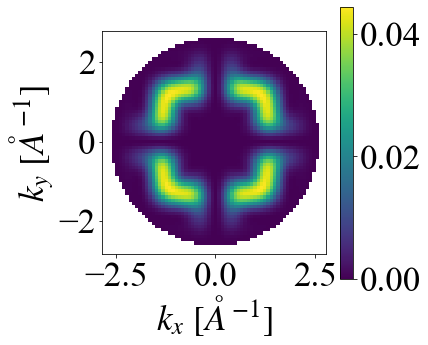}\\
(d)&(e)&(f)\\
\includegraphics[width=0.25\linewidth]{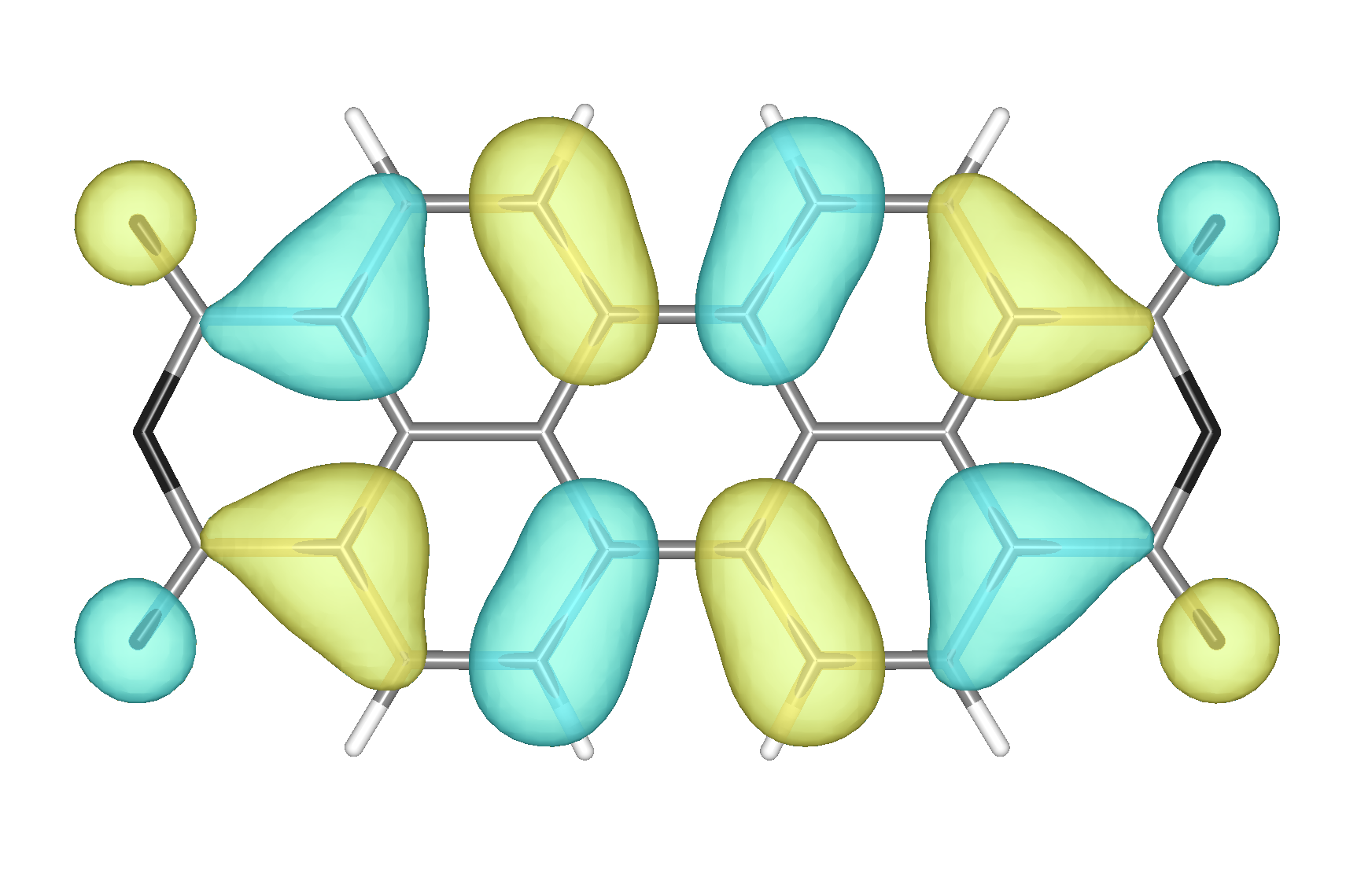}&
\includegraphics[width=0.25\linewidth]{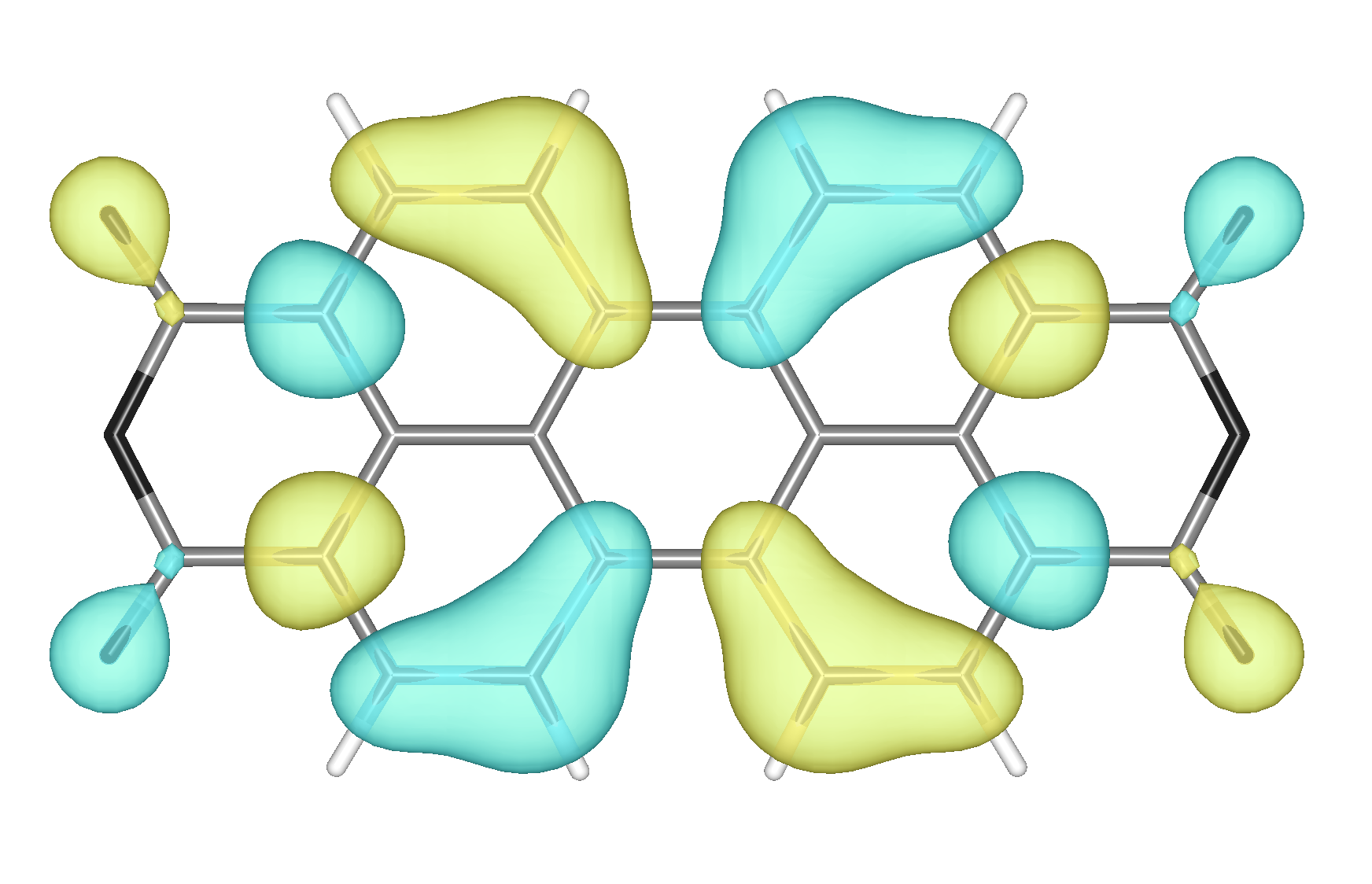}&
\includegraphics[width=0.25\linewidth]{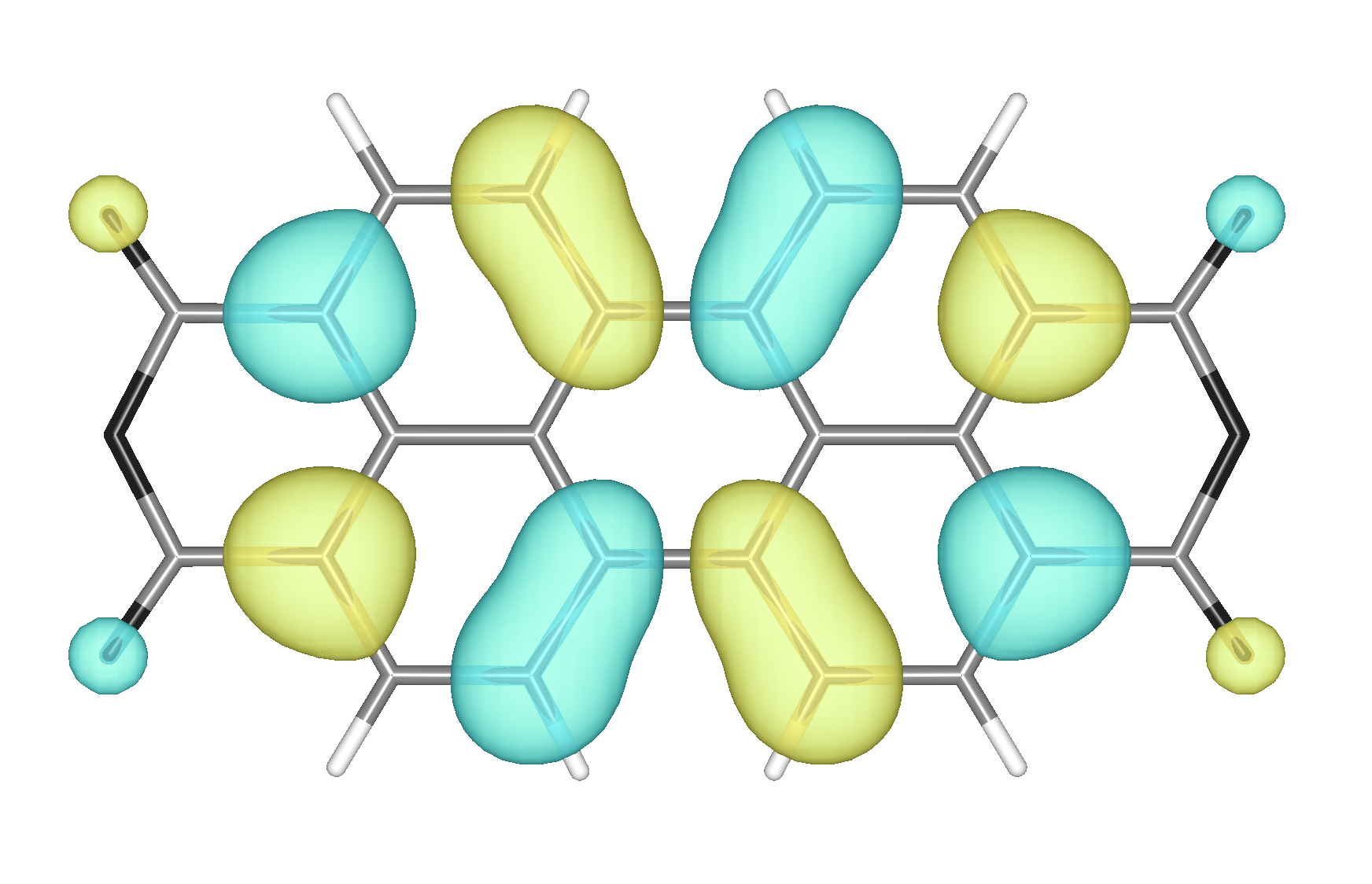}
\end{tabular}
\caption{
(a) Experimental PMM of the HOMO of PTCDA/Ag(110) (original data ware taken from Fig. 2(c) in Ref. \cite{WSS13}). 
(b) The fitted PMM with the least squares approach.
(c) The fitted PMM $\lambda {\bf I}(C)$ with the PhaseLift approach at $\eta'$ = 0.45.
In (b) and (c), the DFT-optimized structure was used.
The plotted data are normalized as ${\bf z}/\|{\bf z}\|_{2}$ for (a) and $\lambda {\bf I}(C)/\|{\bf z}\|_{2}$ for (b) and (c).
(d) The KS-HOMO of isolated PTCDA. 
(e) The estimated orbital with the least squares approach.
(f) The estimated orbital with PhaseLift approach at $\eta'$ = 0.45.
In (d)-(f), the isosurface value is 0.02 a.u.
}
\label{fig:pt}
\end{figure*}

Figures \ref{fig:pt}(b) and \ref{fig:pt}(e) show the fitted PMM and the estimated orbital obtained from the least squares approach. With the estimated orbital, the corner-frame-like features of the experimental PMM [Fig. \ref{fig:pt}(a)] is well reproduced, whereas the features around the $k_x$ = $k_y$ = 0 \AA$^{-1}$ are not reproduced. The normalized distance between the experimental and the fitted PMMs $\|{\bf z} - \lambda {\bf I}({\bf c}{\bf c}^T)\|_{2}/\|{\bf z}\|_2$ is 0.379. Some of the node positions of the estimated orbital [Fig. \ref{fig:pt}(e)] and the KS-HOMO [Fig. \ref{fig:pt}(d)] are different.

\begin{figure}[tb]
\begin{tabular}{ll}
 (a) &\\
\multicolumn{2}{c}{
\includegraphics[width=6cm]{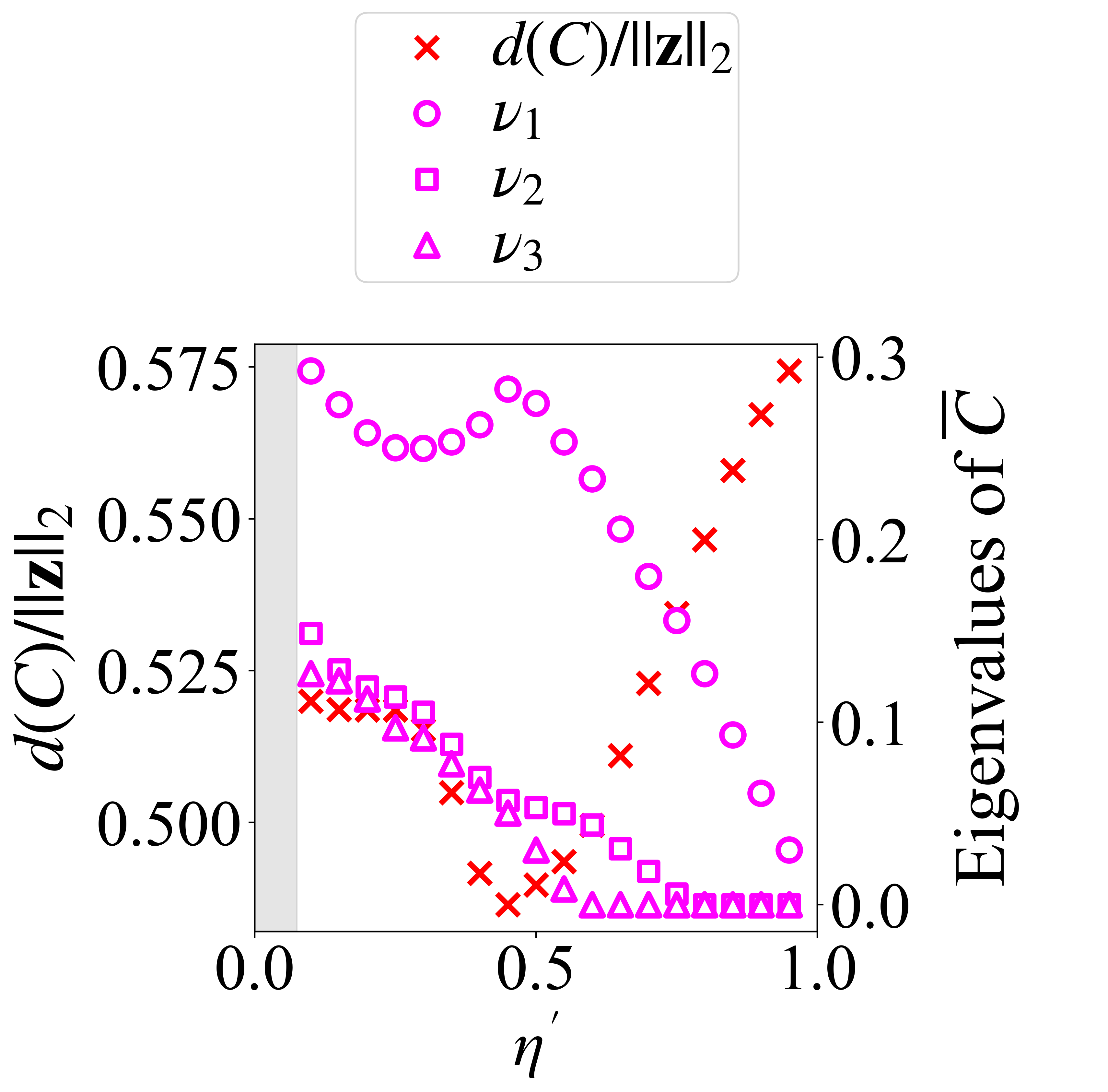}
}\\
(b) &(c)\\
\includegraphics[width=3cm]{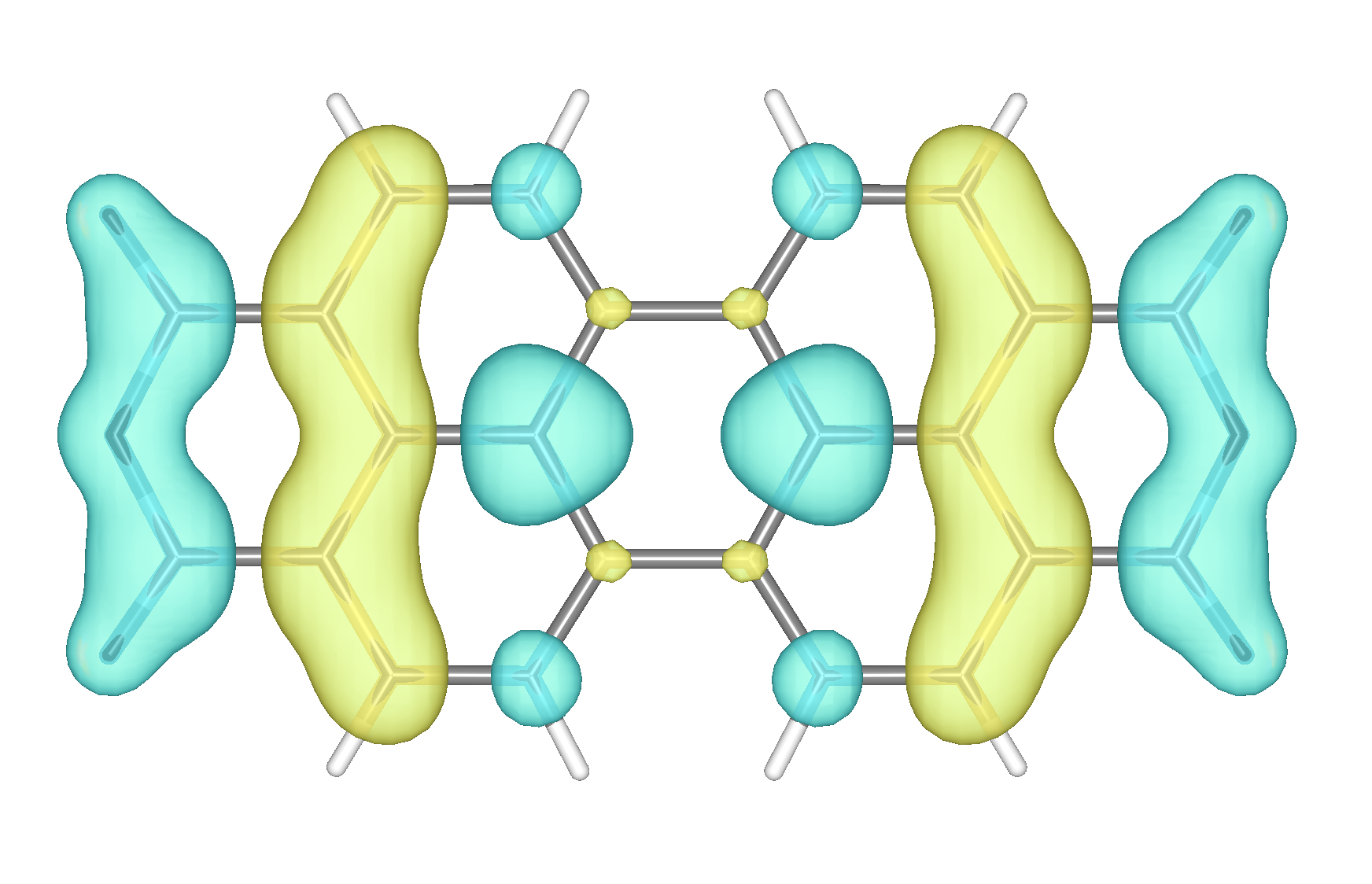}
&\includegraphics[width=3cm]{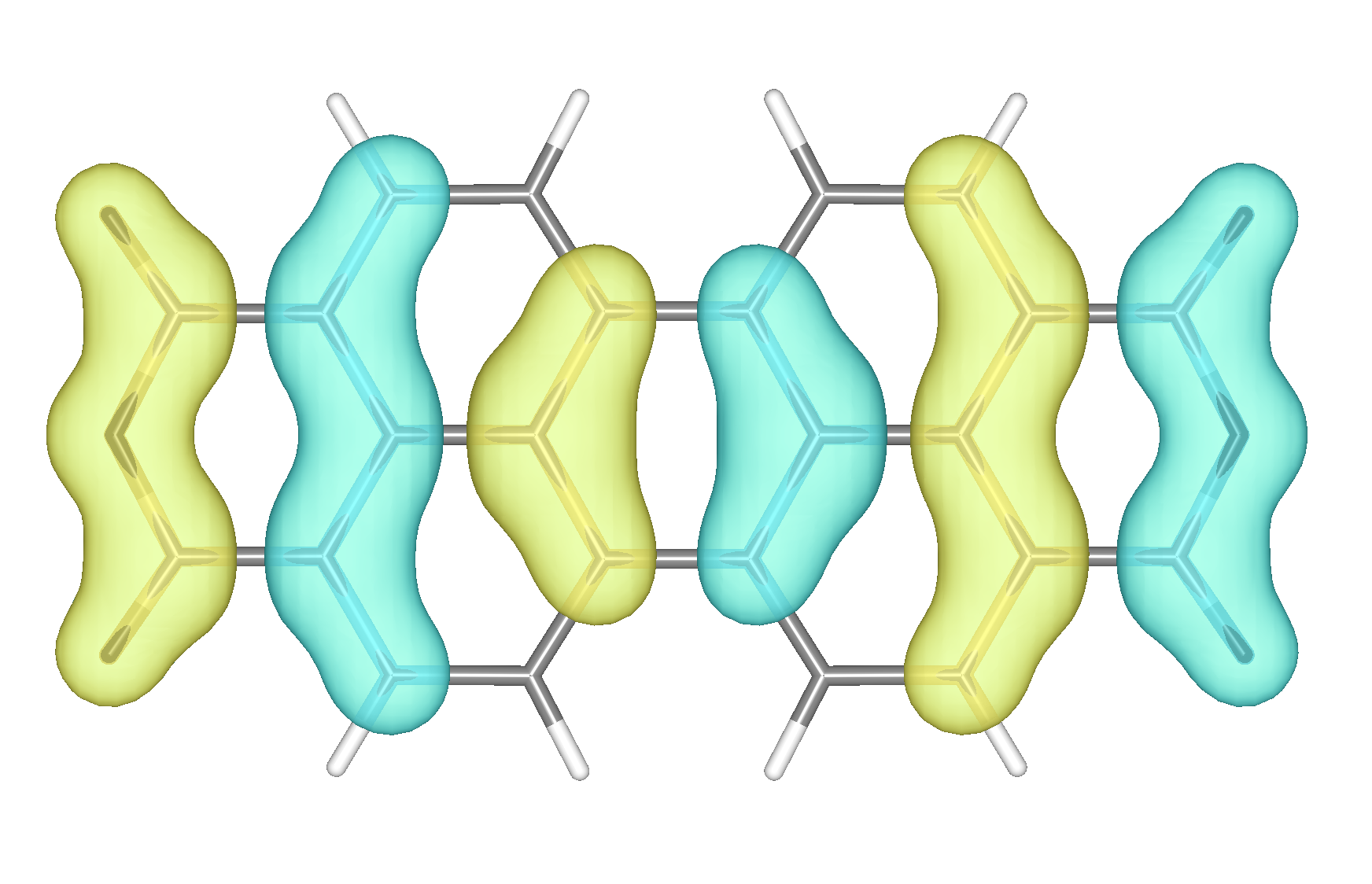}
\end{tabular}
 \caption{
 (a) $d(C)$ and eigenvalues of $\overline{C}$ for PTCDA/Ag(110). Crosses show the distance between the experimental PMM and fitted PMM ($d(C)$, left axis), while open symbols show the largest three eigenvalues of $\overline{C}$ (right axis). In the shaded region, we could not obtain $\overline{C}$. (b) and (c)  represent components with the second and third largest eigenvalues at $\eta'$ = 0.45, respectively.}
\label{fig:pt2}
\end{figure}

In the PhaseLift based approach, $d(C)/\|{\bf z}\|_2$ takes the minimum value 0.486 at $\eta'$ = 0.45 [Fig. \ref{fig:pt2}(a)]. 
Figures \ref{fig:pt}(c) and \ref{fig:pt}(f) show the fitted PMM $\lambda{\bf I}(C)$ and the estimated orbital at $\eta'$ = 0.45. 
With the estimated orbital, the corner-frame-like features of the experimental PMM is well reproduced. The estimated orbital is qualitatively different from the least squares result, but resembles the KS-HOMO [Fig. \ref{fig:pt}(d)] and reconstructed HOMO in Ref. \cite{WLU15}. The origin of the features that were not reproduced by the principal component of $\overline C$ is unclear. The second and the third components at $\eta'$ = 0.45 [shown in Fig. \ref{fig:pt2}(b) and \ref{fig:pt2}(c)] do not coincide with any of the Kohn-Sham orbitals.

\subsection{Simultaneous optimization of molecular structure}
\label{Sec:cc}
One drawback of the present tight-binding based method is that
the precise molecular structure is necessary as an input, which is not always at hand.
In this regard, it is possible to remove the constraints of fixed atomic positions. 
We performed the simultaneous molecular orbital/structural optimization of pentacene using the experimental PMM of pentacene/Ag(110) [Fig. \ref{fig:pen}(a)]. 
We estimated the molecular orbital coefficients using the TRF algorithm and updated the structure using the steepest descent method. We stopped the optimization when the difference of $\left\Vert  {\bf z}-\lambda{\bf I}({\bf c}{\bf c}^T)\right\Vert_{2}/\|{\bf z}\|_2$ is less than $10^{-5}$. We considered possible deformations of the molecule that do not change the D$_{2h}$ symmetry. 

After the PMM-based structural optimization, the normalized distance between the fit and experimental PMM decreased from 0.171 to 0.168. Figure \ref{fig:pen}(d) shows the fitted PMM with the optimized structure. The spatial pattern of the obtained orbital [Fig. \ref{fig:pen}(h)] does not change much from Figs. \ref{fig:pen}(e), (f) and (g) (see Appendix \ref{Sec:appeMO2}). Thus, we can optimize atomic positions as well as molecular orbital coefficients utilizing the information contained in the PMM.

\subsection{POT as molecular structure analysis}
\label{Sec:v0}
Given the above results, it is tempting to assess the possibility to extract
quantitative molecular structural information from PMM data using the present tight-binding
based POT method.
This can be another advantage of using localized atomic orbital basis functions, i.e.,
each of the atomic orbitals used in the fitting has well-defined center at the atomic position in the
molecule and therefore we can expect that the quality of the fit is sensitive to the
atomic positions in the structural model.

Since PMM is essentially a Fourier transform of molecular orbital, accurate absolute values of the photoelectron wavenumber are necessary to set the absolute length scale for structural analysis.

For this purpose, it may be possible to improve the accuracy of the result by introducing a correction to the kinetic energy measured from the vacuum level, that is, if the photoemission from a $\pi$-conjugated molecule is approximately viewed as an escape from a constant potential well, it would be necessary to make a correction to $E_k$ by shifting the origin of the kinetic energy accordingly, corresponding to the inner potential usually introduced as a parameter in the case of solid surfaces. 


In order to check the dependence of the result of the structural optimization on the shift of $E_k \rightarrow E_k+\Delta E_k$, we performed structural optimization with different values of $\Delta E_k$ (= -1 eV, 0 eV, 1 eV, 2 eV, 3 eV).
The calculation with $\Delta E_k$ = 0~eV
is equivalent to the calculation in Sec. \ref{Sec:cc}. 
We assumed that the photoelectron wavevector ${\bm k}'$ inside the potential well is related to that in vacuum by 
\begin{eqnarray}
  {\bf k'} = \sqrt{\frac{2m}{\hbar^2}(E_k + \Delta E_k)}\;{\hat {\bf k}},
\end{eqnarray} 
neglecting the periodic adsorbate structure and assuming the effect of $\Delta E_k$ isotropic.

Resulting representative interatomic distances from PMM-optimized structures are shown in Table \ref{table:structure}. As expected from the property of Fourier transformation, 
the resulting distances decrease as $\Delta E_k$ increases. On the other hand, we obtained almost the same $\left\Vert{\bf z}-\lambda{\bf I}({\bf c}{\bf c}^T)\right\Vert_{2}/\|{\bf z}\|_2$ with these values of $\Delta E_k$. We need additional data, e.g., $h\nu$ dependence of the intensity in order to determine $\Delta E_k$.

The results themselves are promising. From the $\Delta E_k$ dependence, the deviation of the distances corresponding to an uncertainty of $\sim$1~eV in the kinetic energy is  $\sim$0.3\%
in the direction of the long molecular axis and $\sim$3\% in the direction of the short molecular axis.
Also, we can see that the results of the DFT calculation on pentacene/Ag(110) and the experimental value for crystal fall in the range of the values resulting from the PMM optimization, indicating
that the PMM optimization gives values in the plausible range and the PMM data may well contain relevant structural information. 

\begin{table}[hbtp]
  \caption{PMM-optimized (PMM-opt) and DFT-optimized (DFT-opt) structures of pentacene/Ag(110), and experimental (Exp) structure of crystalline pentacene. For the DFT calculations, we used the configurations in Fig.\ref{fig:adsorption}(a) and \ref{fig:adsorption}(b). 
  The experimentally determined crystal structure in fiber-structured thin films is taken from Ref. \cite{pencry}.}
  \label{table:data_type}
  \centering
  \begin{ruledtabular}
    \begin{tabular}{ccccc} 
          method & $\Delta E_k$ (eV) & $l_1$ (\AA) & $l_2$ (\AA) & $\frac{\|\lambda {\bf I}({\bf c}{\bf c}^T)-{\bf z}\|_{2}}{\|{\bf z}\|_{2}}$\\
    \hline
    PMM-opt & -1.0  & 12.36  & 3.03 &0.168\\
    PMM-opt & 0.0 & 12.31  & 3.01 &0.168\\
    PMM-opt & 1.0 & 12.28  & 2.93 &0.168\\
    PMM-opt & 2.0 & 12.26  & 2.84 &0.168\\
    PMM-opt & 3.0 & 12.24  & 2.71 &0.168\\
    DFT-opt(a)& &12.33  & 2.81 &\\
    DFT-opt(b)&  &12.30  & 2.82 &\\
    Exp. (crystal) &&  12.19  & 2.81&\\
    \multicolumn{5}{c}{
    \begin{minipage}{50mm}
      \centering
      \scalebox{0.05}{\includegraphics{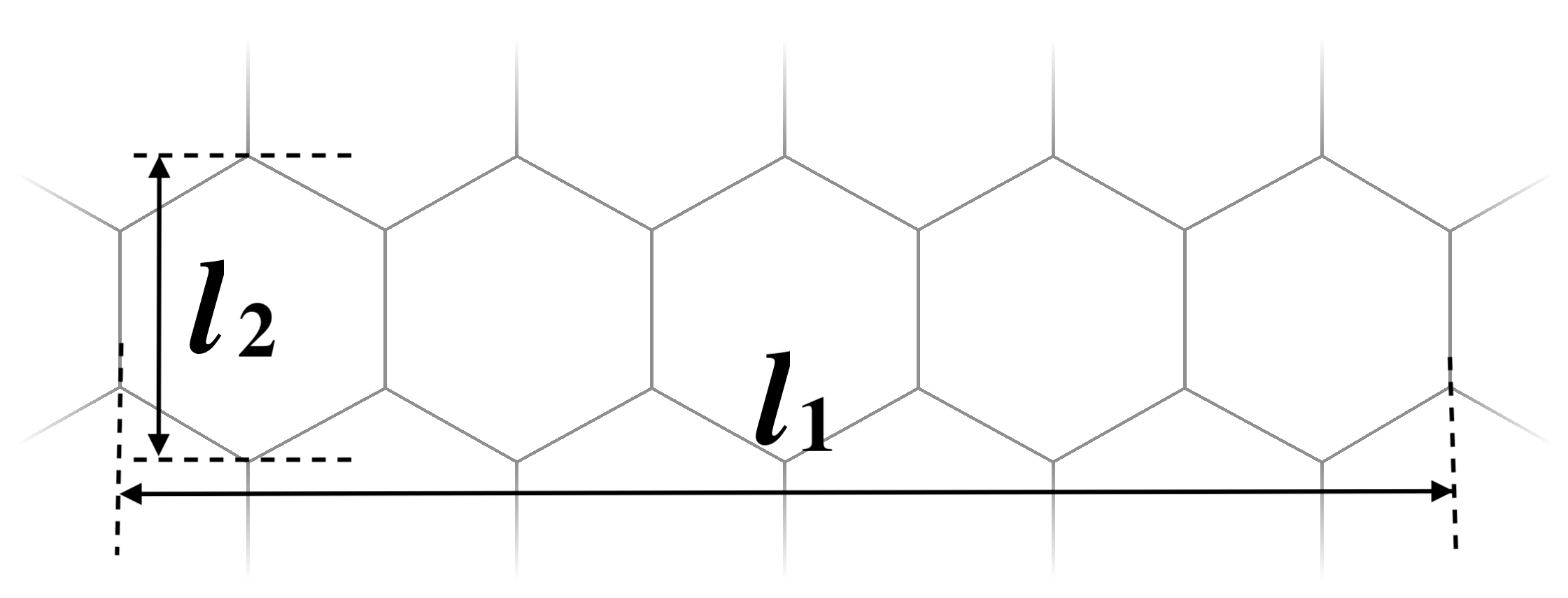}}
    \end{minipage}}
  \end{tabular}
  \end{ruledtabular}
  \label{table:structure}
\end{table}

\section{Conclusion}
In summary, we have presented a photoemission orbital tomography (POT) method based on the LCAO ansatz for the molecular orbitals and showed that it overcomes some of the limitations of the methods based on the Fourier iterative reconstruction.
We successfully reconstructed the 3D HOMOs of pentacene and PTCDA on Ag(110) from 2D PMMs at a single photon energy. 
In the case of PTCDA/Ag(110), we demonstrated that multi-orientational molecular films can be handled, where the PMMs of more than one molecular orientations are superimposed. 
The comparison between the PhaseLift based algorithm and the least squares algorithms showed that the former gives a more plausible orbital in the case of PTCDA/Ag(110) 
by leaving out minor components in the fitting procedure, effectively disregarding the features in the PMM data which are not related to the HOMO of PTCDA or cannot be described by the present simple model of the photoemission process. 
Furthermore, we showed that the method can be used to refine the molecular structure of the pentacene with the experimental PMM.

In principle, information on intermolecular and molecule-substrate orbital hybridization may also be extracted from the PMMs by including orbitals from the surroundings in the basis set when the molecular orbitals remain localized. The present POT method should become a valuable tool for the microscopic understanding of the electronic interactions governing the electronic properties of organic semiconductors.

\begin{acknowledgments}
We would like to thank Manabu Hagiwara, Kazushi Mimura, and Kaori Niki for the fruitful discussions.
We in particular thank Kazushi Mimura for providing us with a sample code to use the PhaseLift algorithm via cvxpy.
M.N. thanks Peter Kr\"{u}ger and Naoya Iwahara for careful reading of the manuscript.
This work was supported by Japan Science and Technology Agency, the Establishment of University Fellowships Towards the Creation of Science Technology Innovation, Grant Number JPMJFS2107.
\end{acknowledgments}

\appendix

\section{Molecular orbital coefficients of HOMO of pentacene}

\renewcommand{\thefigure}{\Alph{section}.\arabic{figure}}
\setcounter{figure}{0}
 
\subsection{Molecular orbital coefficients obtained with different choices of $E_k$} 
\label{Sec:appeMO1}
\begin{figure}[h]
  \centering
   \vspace{1mm}
     \includegraphics[bb= 0 0 651 262, width=0.95\linewidth]{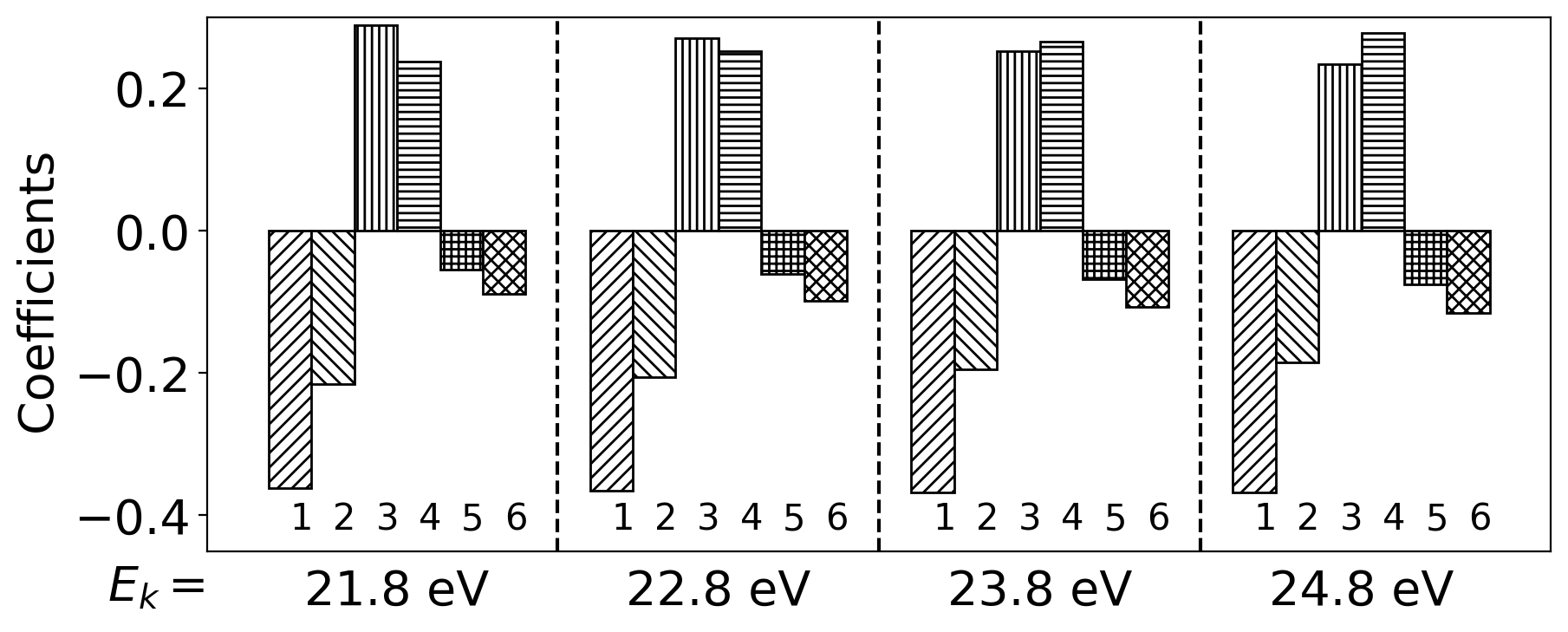}\\
  \hspace*{20mm} 
    \includegraphics[bb= 0 0 2241 855, width=0.5\linewidth]{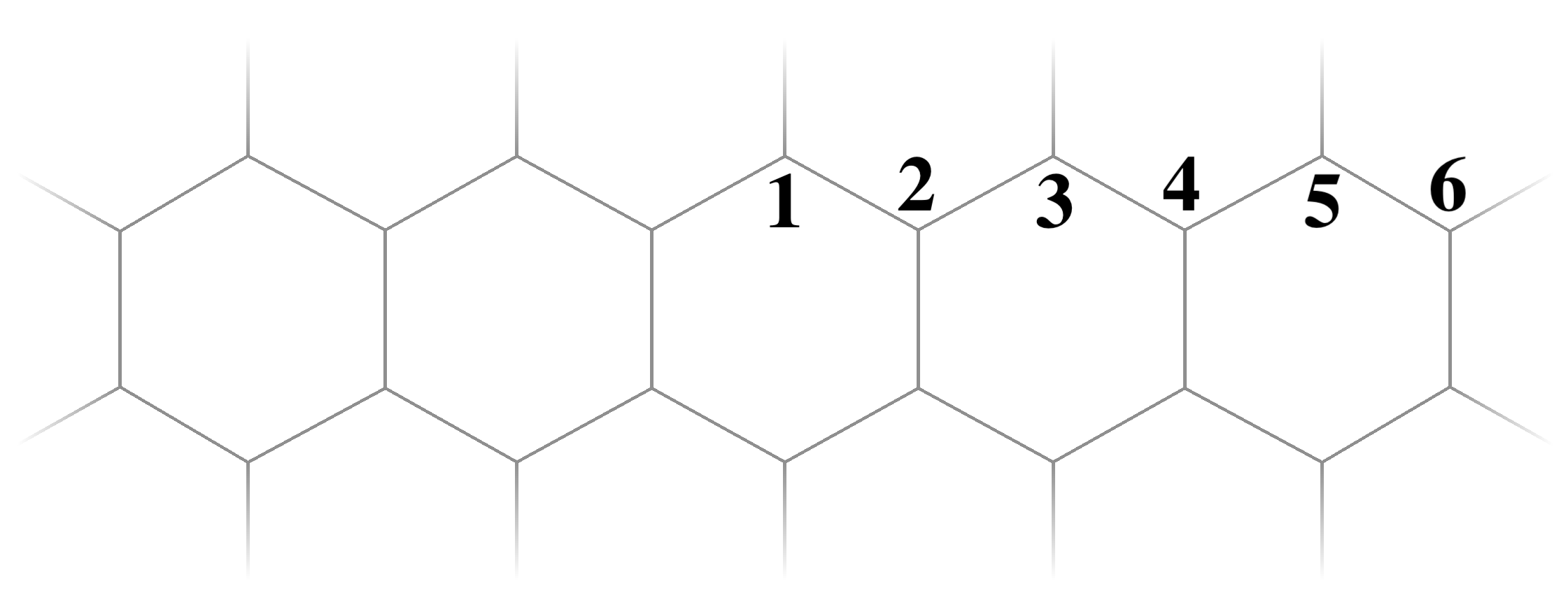}
\caption{$E_k$ dependence of the least squares estimated molecular orbital coefficients using the DFT-optimized structure. Coefficients of the unorthogonalized $2p_z$ orbitals at numbered atomic sites are shown.}
  \label{fig:MOCOEFF1}
  \end{figure}

We show here the least squares estimation of the molecular orbital coefficients of the HOMO of pentacene for different choices of $E_k$ using the same condition described in Sec. \ref{Sec:pen} with $k_z = \sqrt{2mE_{k}/\hbar^2 - (k_x^2 + k_y^2)}$. The results are summarized in Fig. \ref{fig:MOCOEFF1}. 

\subsection{Molecular orbital coefficients  obtained with different structures/methods}
\label{Sec:appeMO2}
\begin{figure}[h]
  \centering
  \includegraphics[bb= 0 0 651 262, width=0.95\linewidth]{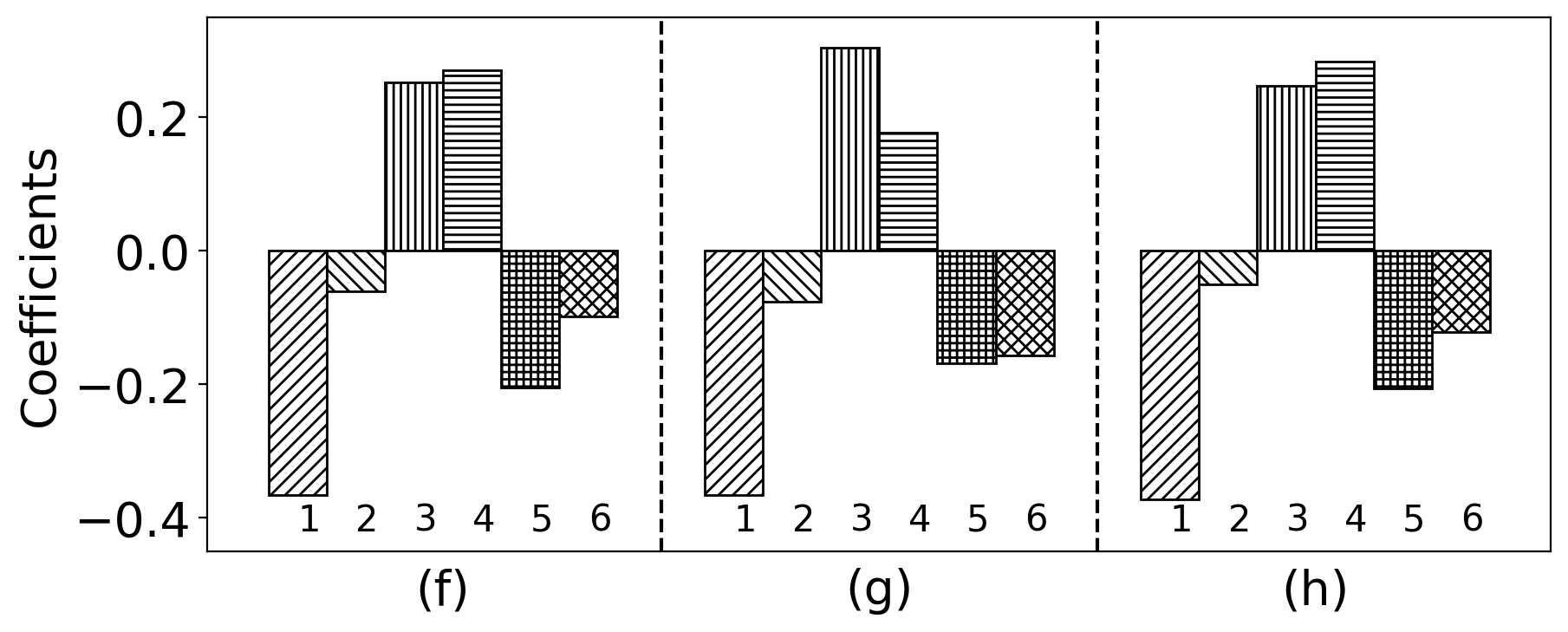}\vspace{1mm}
\hspace*{20mm}
  \includegraphics[bb= 0 0 2241 855, width=0.5\linewidth]{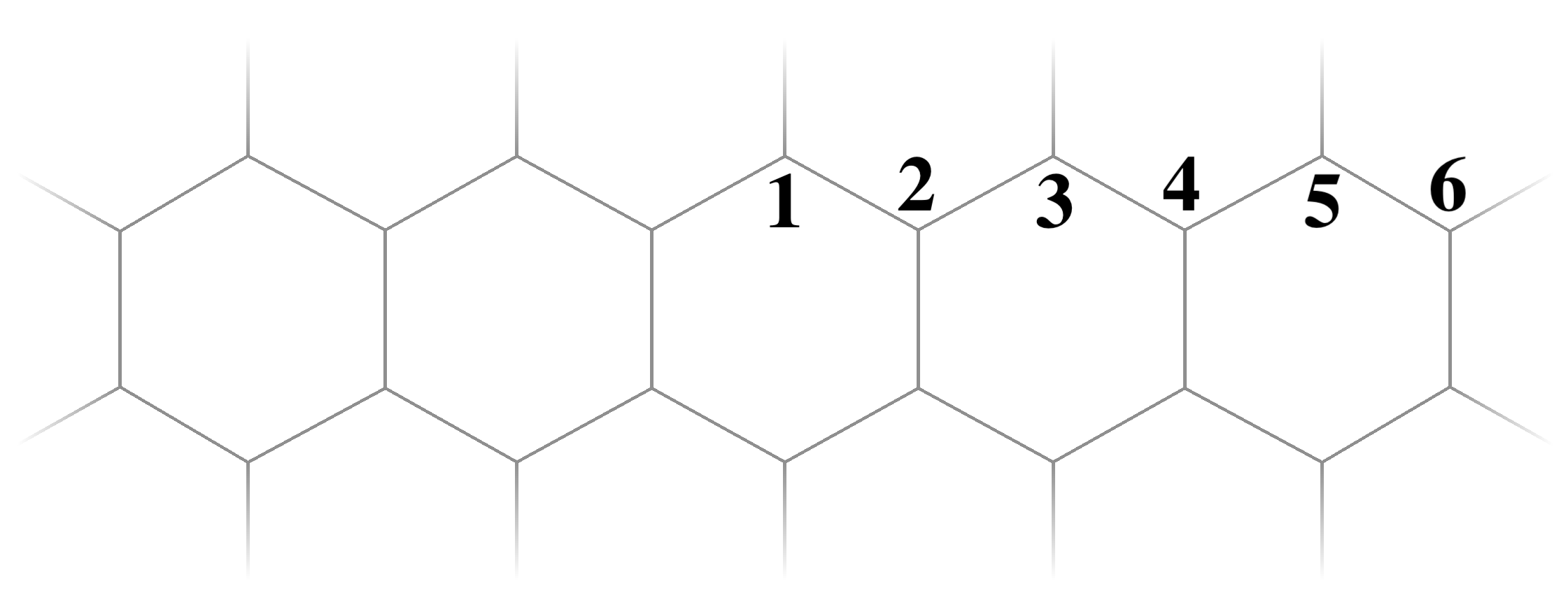}\\
  \caption{Molecular orbital coefficients corresponding to Fig. 3(f), (g) and (h). Coefficients of the unorthogonalized $2p_z$ orbitals at numbered atomic sites are shown.
  }
  \label{fig:MOCOEFF2}
  \end{figure}

For a quantitative comparison of the molecular orbitals in Fig. \ref{fig:pen}(f), (g), and (h), corresponding molecular orbital coefficients are shown in Fig. \ref{fig:MOCOEFF2}. 

\section{Use of larger basis sets in the PMM fitting}
\label{Sec:appe}
In general, it is not possible to improve the orbital reconstruction from 2D PMM by use of split-valence basis sets such as multiple-zeta basis sets.
Suppose that we represent a $\pi$-conjugated molecular orbital as 
\begin{equation}
\psi_i({\bf r})=\sum_n Y_{10}(\widehat{{\bf r}-{\bf R}_n})\sum_{\nu=1}^{N_p}d_{n\nu}R_{1\nu}(|{\bf r}-{\bf R}_n|),
\end{equation}
using $N_p$ radial functions for the $2p_z$ orbital.
Its Fourier transform is
\begin{equation}
{\tilde \psi}_i({\bf k}) = -4\pi iY_{10}(\hat{\bf k})\sum_n e^{-i{\bf k}\cdot{\bf R}_n}
\sum_{\nu=1}^{N_p} d_{n\nu}\tilde{R}_{1\nu}(k) , \label{eq:psi_k}
\end{equation}
where
\begin{equation}
\tilde{R}_{1\nu}(k) = \int_0^\infty dr r^2 j_1(kr) R_{1\nu}(r).\label{eq:pz_int}
\end{equation}
It can be seen from Eq.~\eqref{eq:psi_k} that it is impossible to determine the coefficients $d_{n\nu}$'s at each atomic site, i.e., for the same $n$ and different $\nu$ independently from a data set with fixed $|{\bf k}|=k$. Optimizing $d_{n\nu}$ will be possible with 2D PMMs obtained at several photon energies. 

  \bibliography{ref}

\begin{thebibliography}{45}
\bibitem [{foo()}]{footnote}%
  \BibitemOpen
  \href@noop {} {}\bibinfo {note} {The use of the PhaseLift algorithm in POT
  was suggested to the authors by Manabu Hagiwara and Kazushi Mimura and
  formulation described in this work is similar to that in forthcoming paper
  \cite{OT23}.}\BibitemShut {Stop}%
\end{thebibliography}
  \end{document}